\documentclass[floatfix,twocolumn,showpacs,preprintnumbers,amsmath,amssymb,aip,pop,superscriptaddress,longbibliography]{revtex4-1}

\usepackage{graphicx}
\usepackage{dcolumn}
\usepackage{bm}
\usepackage{float} 
\usepackage{lipsum}
\usepackage{enumitem,amssymb} 
\usepackage{color}
\usepackage[usenames,dvipsnames,svgnames,table]{xcolor}
\usepackage{hyperref}
\hypersetup{
    colorlinks = true,
    citecolor = blue
}
\newlist{todolist}{itemize}{2}
\setlist[todolist]{label=$\square$}
\usepackage{amsmath} 

\usepackage{color}

\usepackage{siunitx}
\usepackage{pifont}

\usepackage{leftidx}
\usepackage{tabulary}
\usepackage{tabularx,booktabs}
\usepackage[capitalise, noabbrev]{cleveref}
\usepackage[utf8]{inputenc}
\usepackage[T1]{fontenc}
\usepackage{csquotes} 
\usepackage[toc,page]{appendix}

\graphicspath{{Figures/}}

\newcommand{\IITJ}{Department of Physics, Indian Institute of Technology Jammu, Jammu, 181221, India}
  
\begin{document}

\title{Elastic properties of three-dimensional Yukawa or dust crystals from molecular dynamics simulations} 

\author{Sandeep Kumar}
\email{sandeepshukla1112@gmail.com} 
\affiliation{\IITJ}


\date{\today}
             
\begin{abstract} 
This paper presents the calculation of elastic properties of three-dimensional Yukawa or dust crystals using molecular dynamics (MD) simulations. The elastic properties are computed by deforming (compressing/expanding) the dust crystals along different directions. The stress and strain of the deformed crystal are used to calculate elastic properties. The bulk modulus, shear modulus, and Poisson's ratio are determined as a function of shielding parameter $\kappa$ and strong coupling parameter $\Gamma$. The bulk and shear modulus values at 0~K temperature are consistent with the previous literature results, while the finite-temperature results are new. The finite-temperature bulk modulus of Yukawa crystals is found to be higher than that of 0~K crystals. The shear modulus of the Yukawa solids decreases nonlinearly near the solid-liquid boundary in the premelting region. The Poisson’s ratio of Yukawa crystals changes sharply at the solid-liquid boundary, emphasizing its potential for identifying phase transitions and assessing incompressibility in Yukawa systems. The bulk and shear moduli calculated in this paper are useful for determining accurate values of sound and shear velocity in Yukawa systems across a wide range of the ($\kappa$, $\Gamma$) parameter space.
\end{abstract} 
\maketitle 
\section{Introduction}
\label{intro}
\setlength{\parindent}{11pt}
Dusty plasmas consist of electrons, ions, neutrals, and immersed dust particles. It is observed in both natural and laboratory settings. In nature, dusty plasmas are observed in Saturn's ring, interstellar clouds, and cometary tails~\cite{Shukla_15}. In the laboratory, it is present in the plasmas of fusion devices, rocket exhaust, and created in experimental devices under controlled conditions~\cite{Chu_94,Chaubey_22}. The dusty plasma offers a model system to study generic phenomena such as self-organization and transport at particle level~\cite{Morfill_09}. Both theoretical and experimental studies have been carried out to study generic phenomena in dusty plasmas, such as crystallization~\cite{Thomas_94,Hamaguchi_97,Maity_19}, single-particle dynamics~\cite{Maity_18,Deshwal_22}, solitons~\cite{Sandeep_17,Bandyopadhyay_08,Donko_20,Prince_23}, shocks~\cite{Sharma_16,Arora_23}, spiral waves~\cite{Sandeep_18_POP, Sandeep_18_PRE}, vortices~\cite{Bailung_20,Gupta_16,Dharodi_20}, Mach cones~\cite{Samsonov_99}, and instability and turbulence~\cite{Wani_24}. 

A typical micron-sized dust particle carries an electronic charge ranging from approximately 10000$e$ to 20000$e$~(where $e$ denotes the charge of an electron) and possesses a mass approximately $10^{13}$ to $10^{14}$ times that of ions~\cite{Wolter_05, Nosenko_04}. Dusty plasma can be modeled by a system of point particles interacting through a pairwise Yukawa potential (screened Coulomb), which is described by the following expression~\cite{Konopka_00}:
\begin{eqnarray} 
{U(r) = \frac{Q^2}{4 \pi \epsilon_0 r } \exp 
(-\frac{r}{\lambda_ D})} \, ,
\end{eqnarray}
where $Q$ is the charge on a dust particle, $r$ is the separation between two dust particles, and $\lambda_ D$ is the Debye length of background plasma. The Yukawa system can be characterized in terms of two dimensionless parameters $\Gamma = {Q^2}/{4 \pi\epsilon_0 a k_B T_d}$ (known as the strong coupling parameter) 
and $\kappa = {a}/{\lambda_D}$ (known as the shielding parameter). Here, $T_d$ and $a$ are the dust temperature and the Wigner-Seitz (WS) radius, respectively. The Yukawa inter-particle interaction is also used to model other systems, including charged colloids~\cite{Palberg,Terao}, electrolytes~\cite{Levin,Lee}, and strongly coupled electron-ion plasmas~\cite{Lyon,Vorberger_12}.

Elastic properties are crucial material characteristics that indicate the stiffness of a material. They are linked to the propagation of waves and their velocities along specific crystallographic directions within a material. Wave propagation and viscoelasticity in dusty plasmas have been extensively investigated through both experiments and simulations~\cite{Morfill_09}. The shear modulus of two-dimensional dusty plasmas has been studied in both experiments and simulations~\cite{Liu_17,Joy_15,Wang_19}. Earlier simulation studies have been carried out to calculate the elastic properties of three-dimensional Yukawa systems~\cite{Robbins_88,Khrapak_20}. However, these properties have not been calculated directly by deforming Yukawa or dust crystals. The elastic properties of Yukawa crystals have been calculated by deforming the crystal using analytical expressions~\cite{Kozhberov_22}. However, this calculation does not account for thermal effects, making it unsuitable for practical applications.

In the present paper the elastic properties of three-dimensional Yukawa or dust crystals are calculated using molecular dynamics (MD) simulations by deforming (compressing/expanding) the crystal along different directions. The elastic properties are calculated for both cold (0~K) and finite-temperatures. The cold calculations are in agreement with the previous literature results, while finite-temperature results provide new insights. The bulk modulus of Yukawa crystals at finite-temperatures is greater than that at~0~K. A prominent result for the finite-temperature shear modulus of Yukawa solids is its nonlinear decrease near the solid-liquid boundary in the premelting region. The present calculations reveal that the Poisson’s ratio of Yukawa crystals undergoes a sharp change at the solid-liquid boundary, highlighting its potential as an indicator of phase transitions and a measure of incompressibility in Yukawa systems. The obtained elastic values are useful for calculating the sound and shear velocity of Yukawa systems across a wide range of the ($\kappa$, $\Gamma$) parameter space.

The paper is organized as follows. Section \ref{mdsim} provides details of MD simulations. In Section \ref{result}, results for the bulk modulus, shear modulus, and Poisson’s ratio are presented as functions of the shielding parameter $\kappa$ and the strong coupling parameter $\Gamma$, which are calculated from three different types of initial particle distributions (random, BCC, and FCC). Section \ref{summary} contains a brief summary of the study.
\section{Simulation Method}
\label{mdsim} 
Following the equilibration (see the details of MD simulation setup and equlibriation in Appendix A), the system is prepared for the calculation of elastic properties under isothermal conditions. First, the cubic box is transformed into a triclinic box and equilibrated again for 450000 time steps. Subsequently, the triclinic box undergoes deformation (expansion and contraction) along the X, Y, Z, XY, XZ, and YZ directions. The distortion along the XY, XZ, and YZ  directions changes the tilt of the triclinic box. After each expansion/contraction, the triclinic box is again equilibrated for 90000 time steps. The resultant changes in the stress of crystal, namely $P_{xx}$, $P_{yy}$, $P_{zz}$, $P_{xy}$, $P_{xz}$, and $P_{yz}$, are used in the calculations of elastic stiffness constants. The details of the formulation of elastic stiffness constants using stress and strain are provided in Appendix B.

The convergence of elastic properties with the magnitude of deformation (expansion/compression) is checked carefully. A deformation of 0.075$L$ (7.5\%~in strain units) is employed in the computations, where $L$ is the simulation box length. The Yukawa or dust crystal remains in the linear region (see Eq.~\ref{eq:voigt_eq} of Appendix B) due to the application of this small strain. The average of stresses is taken at 1000 time steps. The finite size effects on elastic properties are examined by observing changes in the values of bulk modulus, shear modulus, and Poisson’s ratio as a function of the number of particles in the MD simulations. This analysis reveals that the simulations with 10235 particles are sufficient to mitigate such effects. For the reference, no defects were included in the Yukawa crystals.

Most results presented in this paper are being carried out with the random initial distribution of particles. However, for comparison, in some cases, particles are distributed initially at the BCC and FCC lattice sites and then elastic properties are calculated. The bulk and shear modulus results in this paper are presented in normalized units, for which these quantities are normalized by $Q^2/4\pi\varepsilon_0a^4$. Each data point on the plots is obtained by averaging the results obtained from 10 to 20 independent simulations. It was found that the Yukawa liquids required more averaging than the Yukawa solids to obtain statistically accurate elastic properties.

\section{Results}
\label{result}
The purpose of this study is to calculate the elastic properties of three-dimensional Yukawa or dust crystal as a function of $\kappa$ and $\Gamma$. Specifically, the bulk modulus, shear modulus, and Poisson’s ratio are computed to characterize the elastic properties. These properties are also calculated at 0~K temperature to compare the present paper results with the previous literature results. 

\subsection{Bulk Modulus}
The bulk modulus $K$ is a measure of the resistance of a material to an applied bulk compression. In this study, the bulk modulus is calculated using elastic stiffness constants via~\cite{Kittel_18,Robbins_88}
\begin{equation}
K = \frac{C_{11}+2C_{12}}{3} \,.
\label{eq:bulk_eq}
\end{equation}
The details of elastic stiffness constants~$C_{ij}$~formulation are provided in Appendix~B. The bulk modulus calculated using Eq.~\ref{eq:bulk_eq} is a time-independent quantity. 

\begin{figure}[!hbt] 
\centering  
\includegraphics[width=1.0\columnwidth]{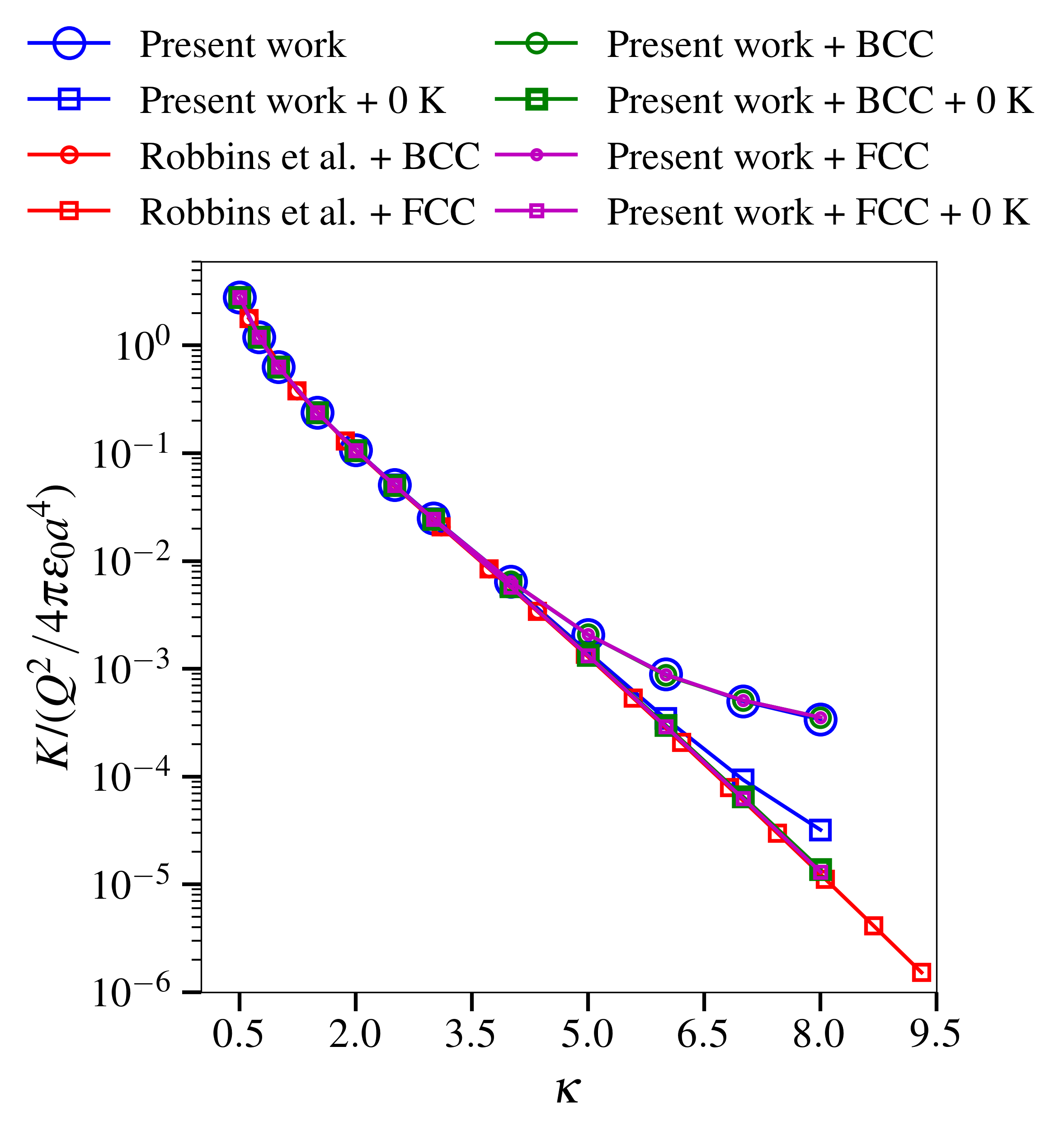}
\caption{Normalized bulk modulus, $K/(Q^2/4\pi\varepsilon_0a^4)$, of Yukawa or dust crystal as a function of shielding parameter $\kappa$. In addition to random distribution (blue points), the bulk modulus is also calculated using BCC (green points) and FCC (magenta points) distributions. The bulk modulus is calculated for both cold (0~K) and finite-temperature cases. Finite-temperature calculations are carried out at the strong coupling parameter $\Gamma$ = $2000$. The reported bulk modulus of~\citet{Robbins_88} (red points) agrees with the cold~(0~K) calculations of present work.}
        
  \label{bm_kp}                    
        \end{figure}

The bulk modulus as a function of screening parameter $\kappa$ is displayed in~Fig.~\ref{bm_kp}. It decreases with an increase in $\kappa$. This reduction occurs because as $\kappa$ increases the Debye length $\lambda_D$ of interaction potential decreases, which makes the compression of the material easier. For large $\kappa$, the Yukawa potential becomes extremely short-range, leading to interactions among particles resembling those in a hard sphere system. Both cold (0~K) and finite-temperature ($\Gamma$ = 2000) calculations are carried out for the study of the bulk modulus. 

The results from the present study are compared with the simulation results of~\citet{Robbins_88} in Fig.~\ref{bm_kp}, and it is found that the bulk modulus values at 0~K temperature with BCC and FCC initial distribution match their values. However, there is a small deviation in $K$ from their values at higher shielding with random distribution. It should be noted that due to the different definition of inter-particle distance in~\citet{Robbins_88}, their shielding parameter ($\lambda$) values have been converted into the form of shielding parameter used in this paper ($\kappa$) by dividing it by $(4\pi/3)^{1/3}$. The bulk modulus of Yukawa systems includes contributions from kinetic and potential components in the stress (see Eq.~\ref{eq:stress_ex} in Appendix~B). At lower $\kappa$ values, the bulk modulus is dominated by interaction contribution, while at higher $\kappa$ values, the kinetic contribution becomes dominant. At higher shielding ($\kappa$ > 3), temperature-driven rigidity takes over the interaction potential-driven rigidity. As a result, the bulk modulus values calculated at finite-temperature are higher than those from the cold (0 K) calculations, and the difference between them increases as the particle shielding increases, see ~Fig.~\ref{bm_kp}. The finite-temperature bulk modulus values obtained from random, BCC, and FCC distribution are closely matched.  
        \begin{figure}[!hbt] 
\centering  
\includegraphics[width=1.0\columnwidth]{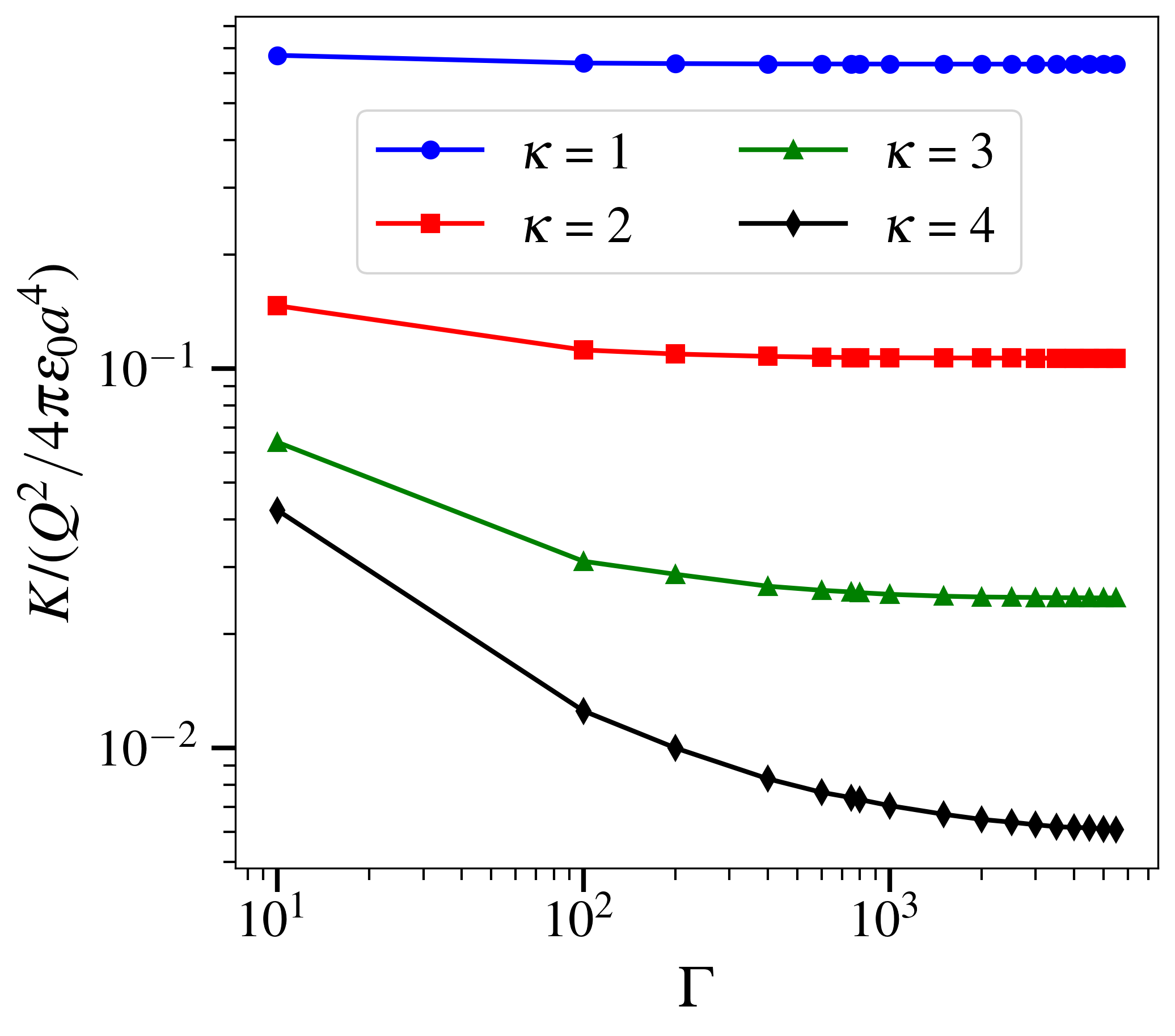}
                   \caption{Normalized bulk modulus, $K/(Q^2/4\pi\varepsilon_0a^4)$, of Yukawa or dust crystal as a function of strong coupling parameter $\Gamma$ for different shielding parameters $\kappa$.}
  \label{bm_gm}	                               
        \end{figure}

The characteristics of bulk modulus with varying strong coupling parameters $\Gamma$ for different $\kappa$ values are shown in~Fig.~\ref{bm_gm}. The bulk modulus decreases as the strong coupling among dust particles increases. In this study, $\Gamma$ is increased by reducing the temperature of the dust particles. Consequently, as strong coupling increases, it becomes easier to compress the particles due to the reduced thermal pressure in the Yukawa crystal. At higher $\Gamma$ values, the rigidity driven by the interaction potential becomes very large in comparison to rigidity driven by temperature, resulting in no further change in bulk modulus with decreasing temperature or increasing strong coupling (see~Fig.~\ref{bm_gm}). The variation in bulk moduli with $\Gamma$ is more pronounced in the high $\kappa$ cases because interaction-driven rigidity is lower in comparison to lower $\kappa$ cases.         

The strongly coupled dusty crystals are extremely soft so that the bulk modulus of solid dusty plasma ranges from $10^{-19}$ to $10^{-16}$~GPa (see Fig.~\ref{bm_gm}), which is much smaller than those of typical solids like metals. For example, the bulk modulus of aluminum at ambient conditions is 79~GPa~\cite{Simmons_65}. 
\subsection{Shear Modulus}
\label{shear_sec}
The shear modulus $G$ is a measure of a material's resistance to an applied shearing deformation. In this study, the shear modulus is calculated using the following relations~\cite{Kittel_18, Robbins_88}:
\begin{equation}
G_1 = C_{44}
\end{equation}
and 
 \begin{equation}
G_2 = \frac{C_{11}-C_{12}}{2} \,.
\end{equation}
Here, $G_1$ and $G_2$ are the shear moduli along (100) and (110) crystallographic planes, respectively. The detailed formulation of the elastic constants $C_{ij}$ is provided in Appendix B. The shear modulus calculated in the present study is a time-independent quantity, commonly referred to as the zero-frequency shear modulus.
\begin{figure}[!hbt] 
\centering  
\includegraphics[width=1.0\columnwidth]{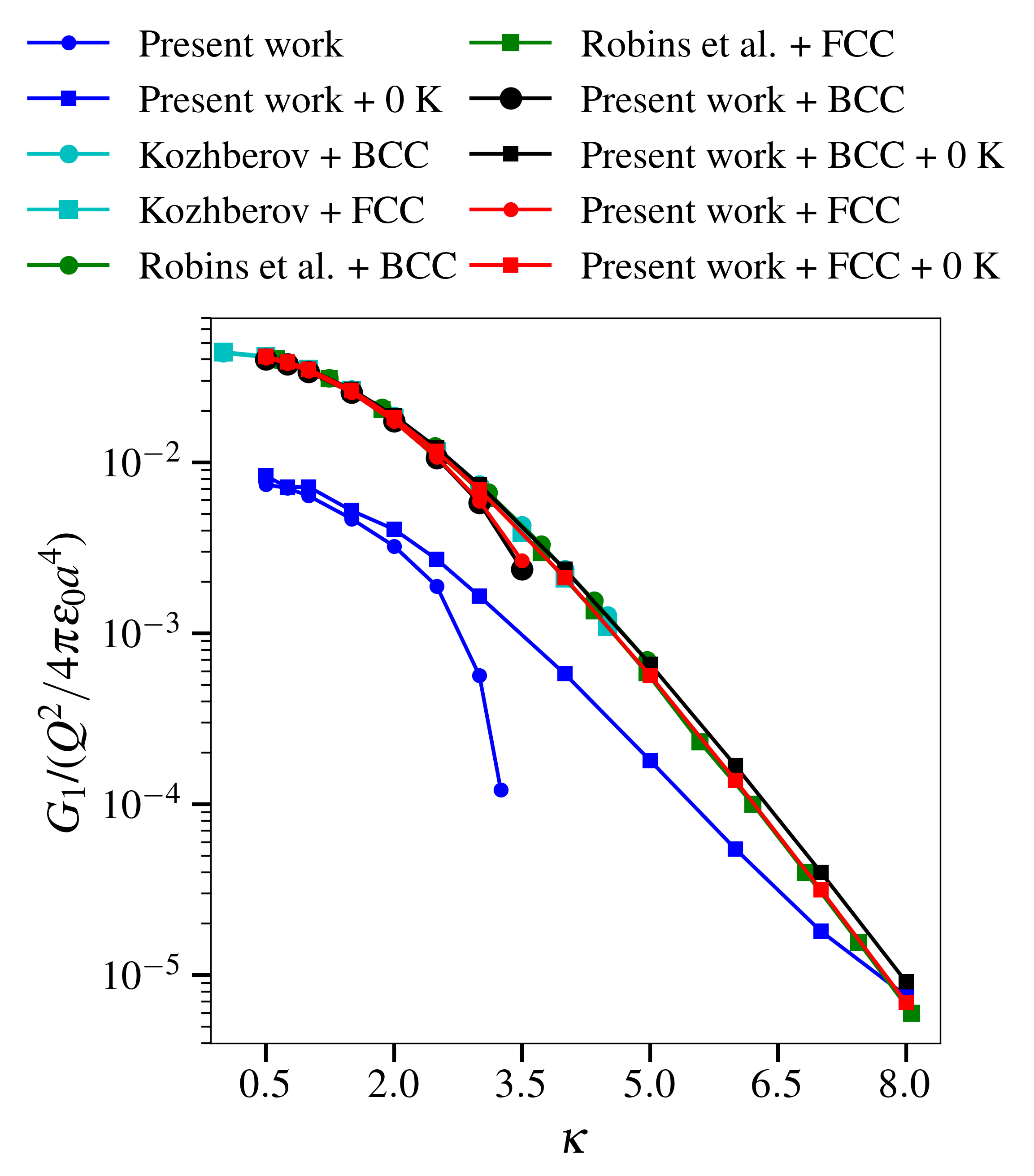}
                   \caption{Normalized shear modulus, $G_1/(Q^2/4\pi\varepsilon_0a^4)$, of Yukawa or dust crystal along the crystallographic plane of (100) as a function of shielding parameter $\kappa$. The shear modulus is calculated for both the cold (0~K) and finite-temperature cases. The strong coupling parameter $\Gamma$ of the Yukawa crystal for the finite-temperature case is 2000. Three types of initial particle distributions—random (blue points), BCC (black points), and FCC (red points)—are used in the calculations of shear modulus. The shear modulus results at 0~K temperature with BCC and FCC distribution are matching with the results of \citet{Robbins_88} (green points) and \citet{Kozhberov_22} (cyan points).}
 \label{sm1_kp}	                       
        \end{figure}  
        
The variation of shear modulus along the crystallographic plane (100), $G_1$, with shielding parameter $\kappa$ is displayed in Fig.~\ref{sm1_kp}. It decreases as shielding parameter $\kappa$ increases, which is a result of the reduction in inter-particle (inter-layer) interactions with the decrease in Debye length $\lambda_D$. At finite temperature, the shear modulus exhibits a nonlinear decrease in the premelting region~\cite{Hamaguchi_97} (for $\Gamma$ = 2000, this occurs between $\kappa$ = 3 to $\kappa$ = 3.5) of the Yukawa system, resembling the shear modulus behavior observed in iron~\cite{Martorell_13}. It exhibits strong shear softening near the melting point; the shear modulus vanishes once the Yukawa crystal melts~\cite{Hamaguchi_97} (for $\kappa$ >  3.5). As with bulk modulus, the shear modulus of Yukawa systems also has both kinetic and potential contributions in the stress~\cite{Donko_08_vis,Joy_15}, see Eq.~\ref{eq:stress_ex} in Appendix~B. At lower $\kappa$ values, the shear modulus is dominated by interaction contribution, while as $\kappa$ increases, the kinetic part also starts contributing. The shear modulus values calculated in the present work are compared with the analytical results of~\citet{Kozhberov_22} and simulation results of~\citet{Robbins_88}, neither of which accounted for temperature in their calculations. For comparison, the shear modulus in this study is calculated with three different types of initial particle distributions in the MD simulations: random, BCC, and FCC. The shear modulus results at 0~K temperature with BCC and FCC distribution match with the results of~\citet{Kozhberov_22} and \citet{Robbins_88}. However, the values with a random distribution are lower than those obtained with BCC and FCC distributions. It should be noted that the shear modulus values at 0~K for higher $\kappa$ values do not vanish, unlike those at finite-temperature in the melted state, where they become zero.

The finite-temperature ($\Gamma$ = 2000) shear modulus values are lower than those calculated at 0~K temperature (see Fig.~\ref{sm1_kp}). At finite-temperatures, the thermal motion of the particles is associated with the kinetic contribution in the shear modulus, which diminishes the shear resistance associated with the interaction contribution. It is unlike the bulk modulus case, where the presence of temperature (thermal pressure) resists the compression of the crystal. As a result, the shear modulus has lower values with the finite-temperatures. The shear modulus values with BCC and FCC distribution are approximately three times higher than those with the random distribution. 
 \begin{figure}[!hbt] 
\centering  
\includegraphics[width=1.0\columnwidth]{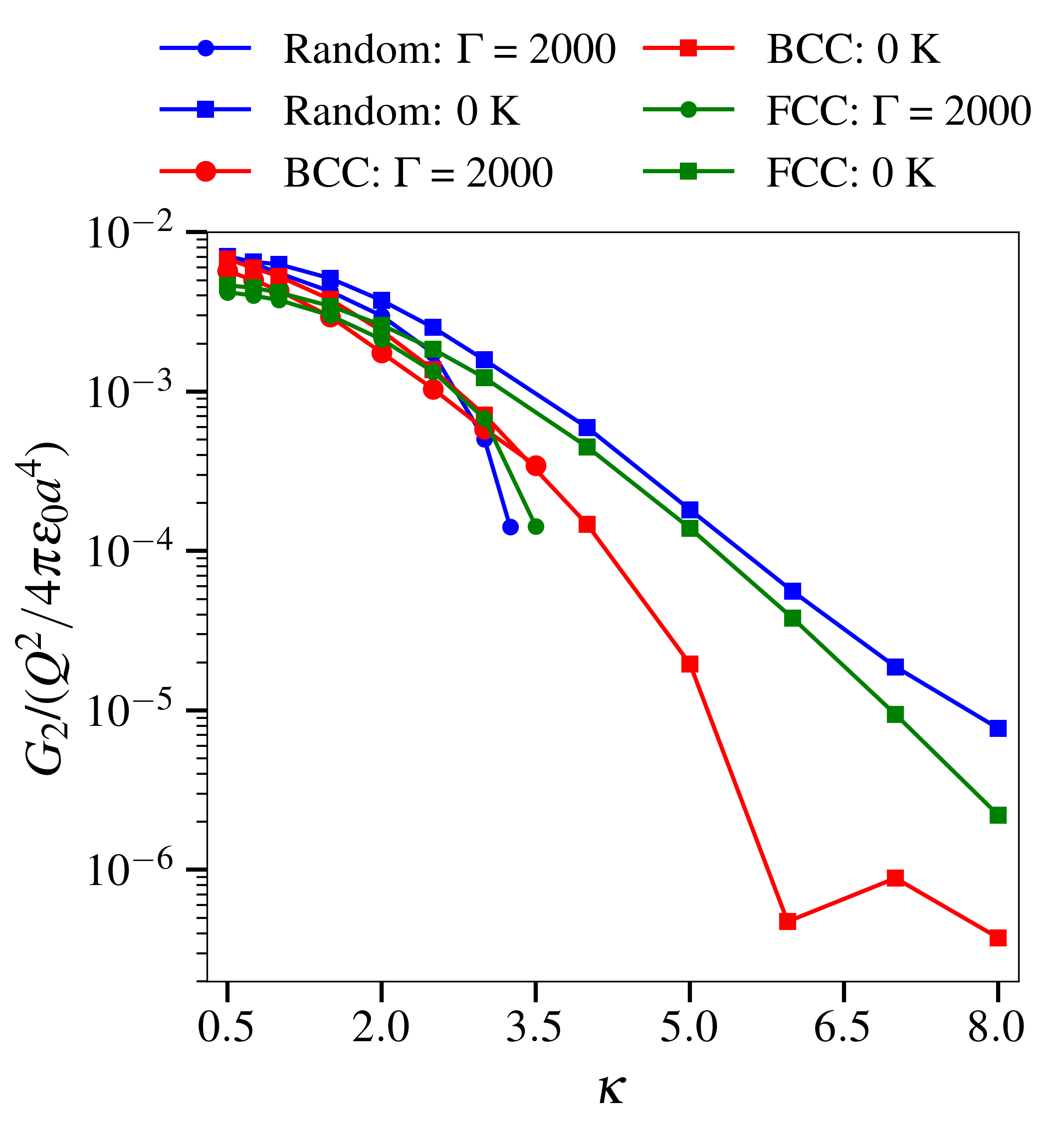}
                   \caption{Normalized shear modulus, $G_2/(Q^2/4\pi\varepsilon_0a^4)$, of Yukawa or dust crystal along the crystallographic plane of (110) as a function of shielding parameter $\kappa$. Three
types of initial particle distributions—random (blue points), BCC (red points), and FCC (green points)—are used in the simulations. Both cold (0~K) and finite-temperature results are shown here.}
        
 \label{sm2_kp}	                       
        \end{figure}       
        \begin{figure}[!hbt] 
\centering  
\includegraphics[width=1.0\columnwidth]{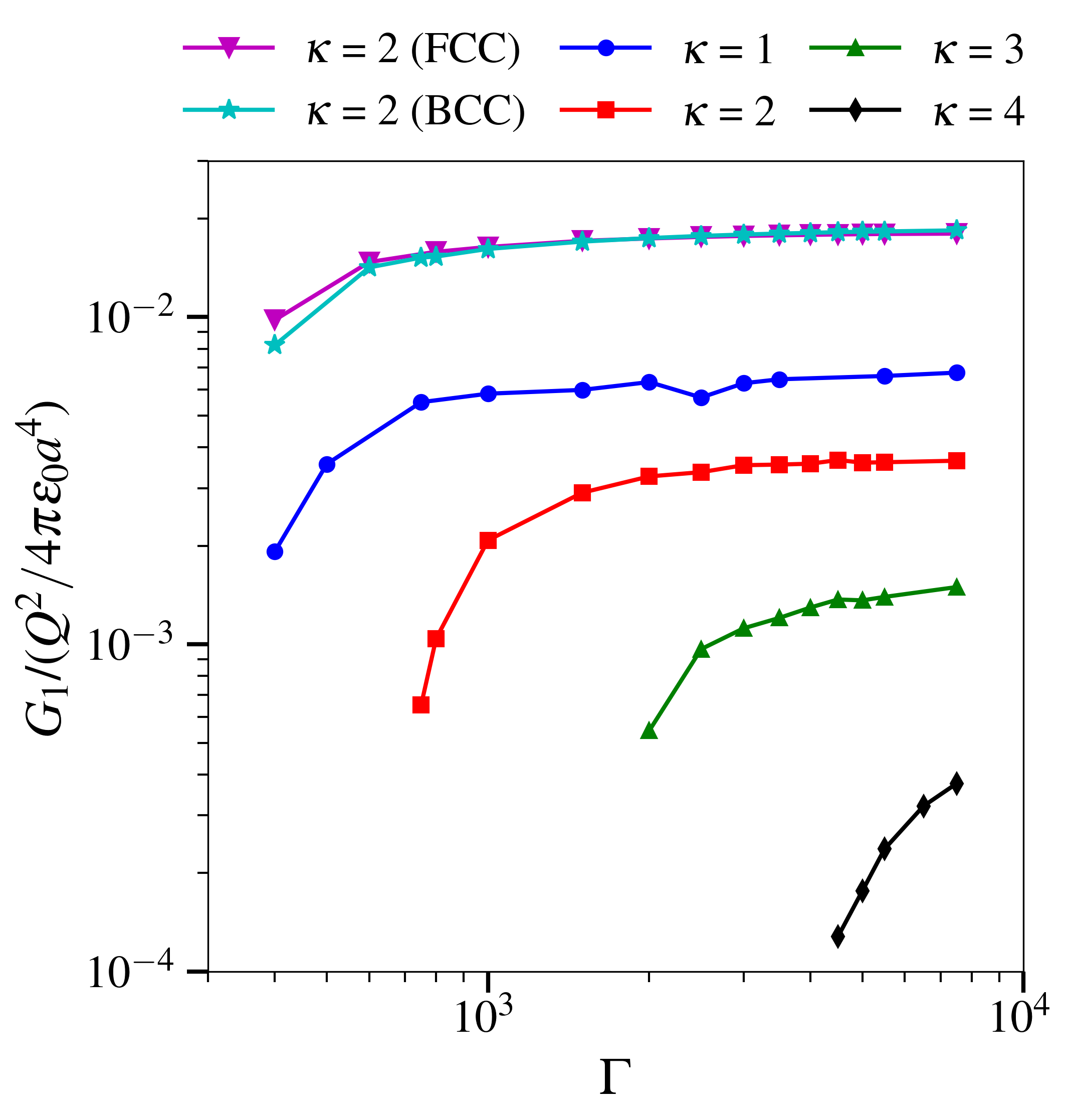}
                   \caption{Normalized shear modulus, $G_1/(Q^2/4\pi\varepsilon_0a^4)$, of Yukawa or dust crystals along (100) crystallographic plane as a function of strong coupling parameter $\Gamma$ for different shielding parameters $\kappa$. For $\kappa$ = 2, in addition to random distribution (red points), the shear modulus is also calculated using BCC (cyan points) and FCC (magenta points) distributions.}
        
  \label{sm1_gm}	                       
        \end{figure}  

The characteristics of shear modulus along the plane (110 ), $G_2$, as a function of $\kappa$ are displayed in Fig.~\ref{sm2_kp}. Similar to $G_1$, it also decreases as the shielding parameter $\kappa$ increases. However, the difference between the shear modulus values calculated from random, BCC, and FCC distribution is not as much as it is in the $G_1$ case. This occurs because particle arrangement along the (110) crystallographic plane differs from that along the (100) plane, causing the shear stress to distribute along the (110) plane differently. The values of $G_1$ and $G_2$ (see Figs. \ref{sm1_kp} and \ref{sm2_kp}) differ for the same reason. 

The shear modulus along the crystallographic plane (100), $G_1$, as a function of strong coupling parameter $\Gamma$ for different $\kappa$ values is plotted in~Fig.~\ref{sm1_gm}.  At lower temperatures (higher $\Gamma$), the shear stress is dominated by the interaction contribution, while as temperature increases ($\Gamma$ decreases), the kinetic part also starts contributing (see Eq.~\ref{eq:stress_ex} in Appendix~B). At very high $\Gamma$ values, the temperature-driven shear modulus becomes small and is dominated by potential-driven shear modulus. Consequently, variation in shear modulus with strong coupling becomes negligible (see~Fig.~\ref{sm1_gm}). A prominent feature of~Fig.~\ref{sm1_gm} is the sharp decrease in the shear modulus of Yukawa crystals near the solid-liquid boundary in the premelting region. This characteristic has also been observed for iron at premelting conditions~\cite{Martorell_13}. The zero-frequency shear modulus becomes zero once Yukawa crystals melt. Therefore, the present work does not report the shear modulus of Yukawa systems in the liquid state.


Furthermore, the shear modulus values obtained with the BCC and FCC distribution are greater than those obtained from the random distribution; see $\kappa =2 $ curves over $\Gamma$ range obtained from random, BCC, and FCC distribution in~Fig.~\ref{sm1_gm}. Nevertheless, the overall characteristics of the curves are similar. The shear modulus values depend on the crystallographic plane along which it is calculated, see Figs.~\ref{sm1_kp} and~\ref{sm2_kp}. In the solid phase with a random initial distribution, particles are oriented in different directions. As a result, averaging over these orientations leads to a lower shear modulus. This is also the reason for the lower shear modulus values observed for the random distribution in Fig.~\ref{sm1_kp}. \citet{Robbins_88} have also discussed the effect of orientation on the shear modulus values in the case of polycrystalline materials, where variations in particle alignment result in differing shear responses. In Fig.~\ref{sm1_gm}, the $\Gamma$ value at which premelting occurs varies for random, BCC, and FCC distributions; the Hysteresis effect could be the reason for this difference~\cite{Morris_94,Zhang_12}.

Similar to the bulk modulus, the shear modulus of Yukawa crystal is much smaller than that of typical solids, which varies from $10^{-20}$ to $10^{-18}$~GPa~(see~Fig.~\ref{sm1_gm}). In comparison, the shear modulus for aluminum under ambient conditions is 25.5 GPa~\cite{Archer_72}.
\subsection{Poisson's Ratio}
\begin{figure}[!hbt] 
\centering  
\includegraphics[width=1.0\columnwidth]{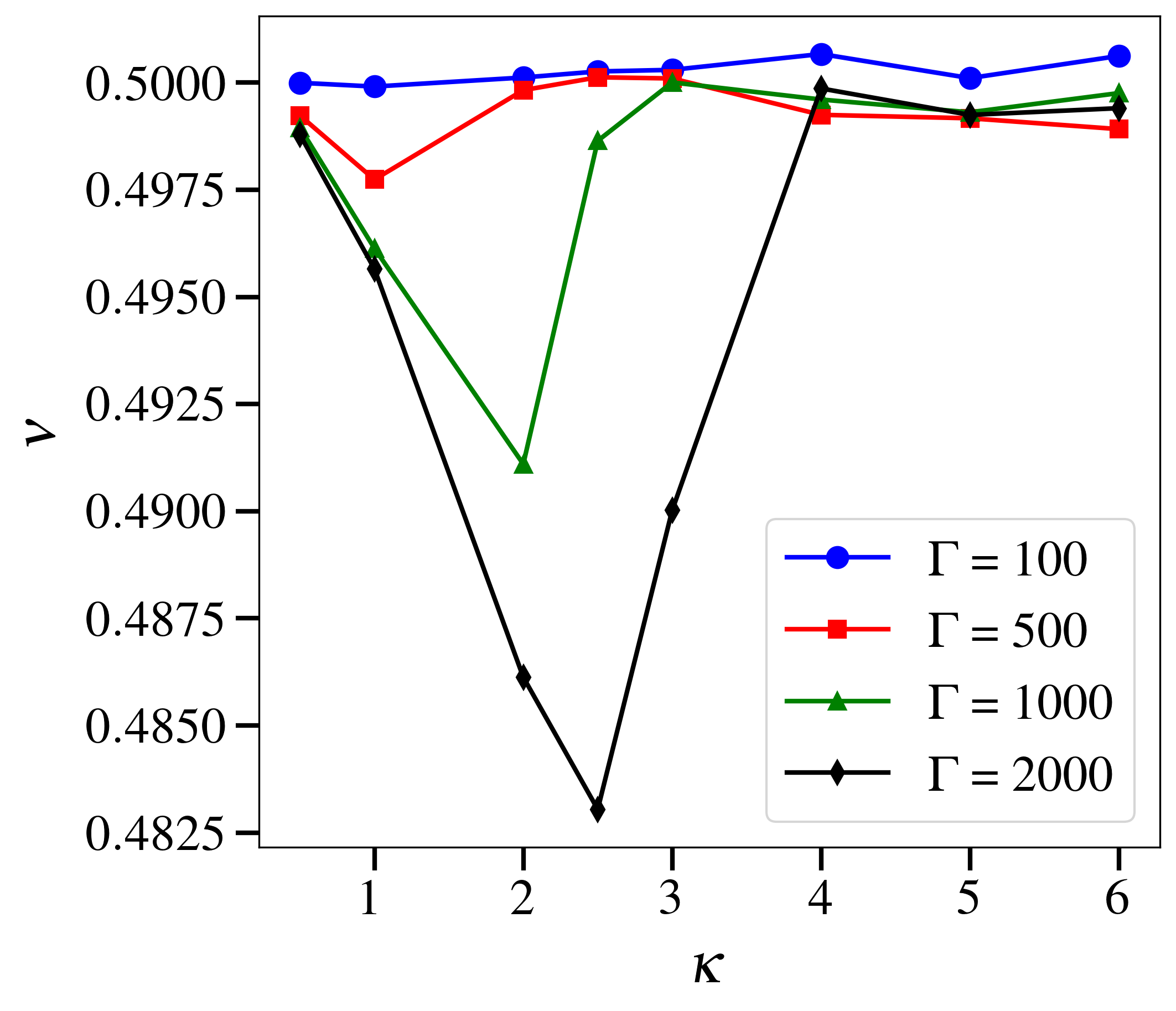}
                   \caption{Poisson's ratio $\nu$ of Yukawa or dust crystals as a function of shielding parameter $\kappa$ for different strong coupling parameters $\Gamma$.}

 \label{pr_kp}       	                       
        \end{figure}

The Poisson's ratio $\nu$ quantifies the deformation (compression/expansion) of a material in directions perpendicular to a specific direction of loading. It is defined as the negative ratio of transverse strain to axial strain. In the present study, Poisson's ratio is calculated using the following relation~\cite{Mott_09}:
 \begin{equation}
\nu = \frac{1}{1 + \frac{C_{11}}{C_{12}}} \,.
\end{equation}
In the above expression, $C_{11}$ and $C_{12}$ are the elastic stiffness constants. The characteristics of Poisson's ratio as a function of $\kappa$ for different $\Gamma$ values are displayed in~Fig.~\ref{pr_kp}. Depending on $\kappa$ and $\Gamma$ values, the Yukawa system transitions into a solid (BCC or FCC) or liquid state, as discussed in ref.~\cite{Hamaguchi_97}. In the solid phase, initially, Poisson's ratio decreases with an increase in $\kappa$ because the transverse strain decreases with the reduction in Debye length of potential $\lambda_D$; however, with further increase in shielding parameter, it increases. This behavior occurs due to structural changes in the Yukawa system, resulting in particles interacting differently and stress getting distributed across the crystal differently. In the liquid phase of the Yukawa system, the Poisson's ratio does not change with $\kappa$ (see Fig.~\ref{pr_kp}) due to the uniform distribution of stress across the Yukawa liquid. The value of $\nu$ in the liquid state is approximately 0.5.  

\begin{figure}[!hbt] 
\centering  
\includegraphics[width=1.0\columnwidth]{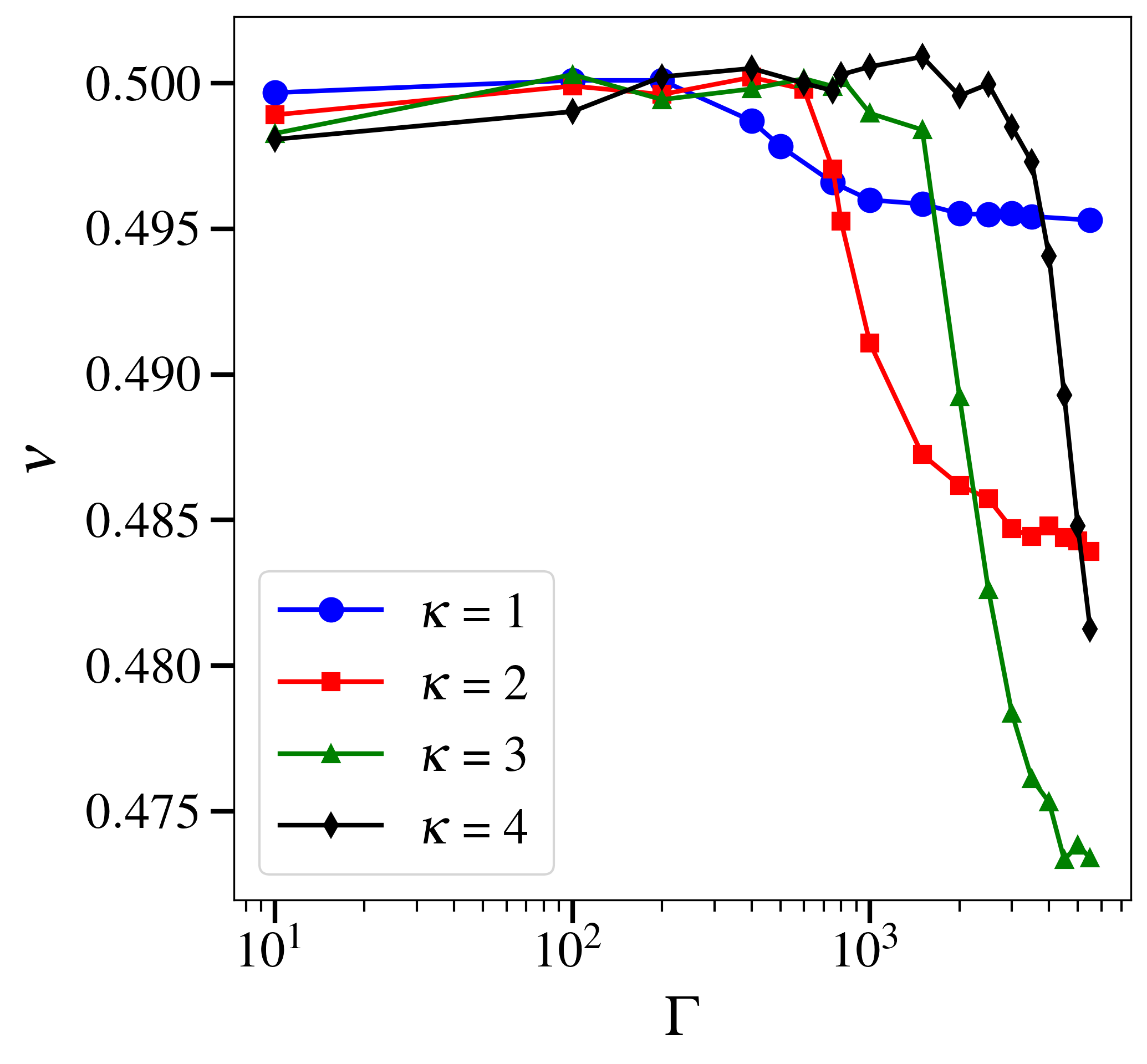}
                   \caption{Poisson's ratio $\nu$ of Yukawa or dust crystals as a function of strong coupling parameter $\Gamma$ for different shielding parameters $\kappa$.}
        
 \label{pr_gm}                       
        \end{figure}

The variation of Poisson's ratio $\nu$ with strong coupling for different $\kappa$ values is shown in~Fig.~\ref{pr_gm}. In the liquid phase, Poisson's ratio does not change with increasing $\Gamma$. However, at the solid-liquid boundary, it changes sharply. In the solid phase, Poisson's ratio decreases with an increasing strong coupling, which occurs due to reduced strain in the transverse direction caused by less thermal motion (thermal pressure) of the particles at low temperatures. Fig.~\ref{pr_kp} and Fig.~\ref{pr_gm} demonstrate that $\nu$ could be used to identify solid-liquid boundaries in Yukawa systems or dusty plasmas.         

For Yukawa or dust crystals, the bulk modulus is significantly larger than the shear modulus, so they can be regarded as effectively incompressible, as it is easier to change shape than compress. Soft materials exhibiting similar characteristics are also considered incompressible in nature~\cite{Jastrzebski_76}. A perfectly incompressible isotropic material has a Poisson’s ratio of exactly 0.5, while it is approximately 0.5 for dusty plasmas. The incompressible nature of dusty plasma changes with variations in $\kappa$ and $\Gamma$, see~Fig.~\ref{pr_kp}~and~Fig.~\ref{pr_gm}. Therefore, Poisson's ratio could be used to characterize the incompressible nature of Yukawa systems or dusty plasmas. For comparison, the Poisson’s ratio of natural rubber and aluminum is 0.499 and 0.33, respectively~\cite{Mott_09}.
\section{Summary} 
\label{summary}
In this study, the bulk modulus, shear modulus, and Poisson's ratio of a Yukawa or dust crystal have been calculated using MD simulations. These elastic properties have been computed by deforming (compressing/expanding) the dust crystal along different directions. The characteristics of the elastic properties as a function of shielding parameter and strong coupling parameter have been presented. The bulk and shear modulus results at 0~K temperature match well with previous literature data, while the results presented for finite-temperatures are new. It has been found that the bulk modulus of Yukawa crystals at finite-temperatures is higher than that at 0~K. The shear modulus of the Yukawa or dust crystal depends on the crystallographic plane along which it is calculated. Additionally, it has been found that the shear modulus values depend on the initial distribution of particles in the MD simulations, whether random, BCC, or FCC. This occurs because particles are oriented in various directions with a random initial distribution and averaging over these orientations leads to different shear modulus values.

The shear modulus decreases nonlinearly near the solid-liquid boundary in the premelting region, which is a prominent result. It exhibits strong shear softening near the melting point. Previous studies on the shear modulus have not reported such a decrease in their calculations. The Poisson's ratio of the Yukawa crystal also changes sharply at the solid-liquid boundary, highlighting its potential use in identifying phase transitions in Yukawa systems. In the present work, it has been found that the bulk modulus is much larger than the shear modulus, which effectively renders Yukawa or dust crystal incompressible. This is due to the fact that it is easier to change shape than to compress a Yukawa or dust crystal. It would be interesting to study the time-dependent shear modulus (shear relaxation modulus) of three-dimensional Yukawa liquids as a function of $\kappa$ and $\Gamma$, particularly in the visco-elastic range. This is a subject left for future work.

The calculated elastic properties can be used to determine accurate values of the longitudinal sound wave velocity (=$\sqrt{(K + \frac{4}{3}G)/\rho}$) and transverse shear wave velocity (=$\sqrt{G/\rho}$) in three-dimensional Yukawa or dust crystals; calculating the characteristic mode velocities of the Yukawa systems using elastic properties is more computationally efficient than employing the wave dispersion\cite{Donko_08,Sandeep_23}(current spectrum) method. The nonlinear decrease of the shear modulus in the premelting region could serve as a foundation for new research explorations in Yukawa or dust crystals. The results reported in the present paper can also apply to other systems, such as charged colloids, electrolytes, and strongly coupled electron-ion plasmas.

\begin{acknowledgments} 
The author thanks Dr. Sanat Kumar Tiwari, Mr. Rauoof Wani, Ms. Farida Batool, Dr. Timothy Callow (CASUS, Görlitz, Germany), Dr. Attila Cangi (CASUS, Görlitz, Germany), and Dr. Svetoslav Nikolov (SNL, Albuquerque, USA) for helpful discussions. The computations were performed on the Agastya Cluster of IIT Jammu, on a Bull Cluster at the Center for Information Services and High-Performance Computing (ZIH) at Technische Universit\"at Dresden, on the cluster Hemera at Helmholtz-Zentrum Dresden-Rossendorf (HZDR), and on TACC Frontera at the University of Texas at Austin. This work is supported by Anusandhan National Research Foundation (ANRF), New Delhi, under the National Postdoctoral Fellowship (NPDF) scheme, project No. NPDF/2023/003366.
\end{acknowledgments}  

\section*{Data Availability}
The data that support the findings of this study are available from the author upon reasonable request.

\appendix
\section*{Appendix A: MD simulation setup and equilibriation}
\label{sec_app_elas_MD}
\setcounter{equation}{0}
\addcontentsline{toc}{section}{Appendix}
\renewcommand{\theequation}{A\arabic{equation}}
A three-dimensional cubic box containing point particles is created to study the elastic properties of dust crystals. The Yukawa interaction potential is taken among the dust particles, which mimics the screening from background electrons and ions. The Large-scale Atomic/Molecular Massively Parallel Simulator (LAMMPS)~\cite{Thompson_22} is used for MD simulations. The simulation box contains 10235 dust particles within dimensions of 35$a$$\times$35$a$$\times$35$a$ (ranging from 0 to 35$a$) along the X, Y, and Z directions, respectively, where $a$ is the average inter-particle distance (Wigner-Seitz radius). Experimental parameters~\cite{Bailung_20} are employed in the simulations: $Q$ = 20000$e$ (where $e$ denotes the charge of an electron), dust mass m = 1.7$\times$$10^{-13}$~Kg, and average inter-particle distance $a$ = 6$\times$$10^{-4}$ m. The average inter-particle distance sets the number density of dust particles $n$ = 1.1$\times$$10^{9}$~$\text{m}^{-3}$, corresponding to a mass density $\rho$ of 1.87$\times$$10^{-4}$~$\text{Kg}$~$\text{m}^{-3}$. The characteristic frequency of the dust particles $\omega_{pd}$ = $(nQ^2/\epsilon_0m)^{1/2}$ is 86.81 s$^{-1}$. In the simulations, the value of shielding parameter $\kappa$ is increased by decreasing the screening length $\lambda_D$ of the Yukawa interaction, and the value of strong coupling parameter $\Gamma$ is increased by decreasing the temperature of dust particles. 

The dust particles are distributed randomly or on BCC/FCC lattice sites in the simulation box and equilibrated at a given temperature using Langevin dynamics~\cite{Schneider_78}, which reads as
 \begin{equation}
m\ddot{\mathbf{r}_\mathrm{i}} = -\sum_{j}\mathbf{\nabla} U_{ij} + \mathbf{F}_{f} +\mathbf{F}_{r} \, ,
\end{equation}
where $U_{ij}$ is interaction potential, $\mathbf{F}_{f}$ is the frictional force on the particles, and $\mathbf{F}_{r}$ is the random force (kicks) on the particles. The frictional force $\mathbf{F}_f$ arises from the relative velocity $\mathbf{v}$ between the Yukawa (dust) particles and the background particles, and is expressed as~\cite{Plimpton19951, Schwabe_2013, Sandeep_18_PRE}:
\begin{eqnarray}
\mathbf{F}_{f} = -m\nu \mathbf{v}.
\end{eqnarray}
Here, $\nu$ is the damping coefficient. The force $\mathbf{F}_{r}$ represents the random kicks experienced by Yukawa particles due to collisions with background species. It is given by
\begin{eqnarray}
\mathbf{F}_{r} \propto \sqrt{\frac{k_B T_b m \nu}{dt}}\,,
\end{eqnarray}
which depends on the time step of simulation $dt$ and the background particle temperature $T_b$.

In this study, the simulation time step of 100~$\mu$s~($\approx$0.009$\omega_{pd}^{-1}$) is chosen, which ensures a fine discretization along the temporal domain and good resolution of the underlying dust kinetics. A damping parameter of 10 s$^{-1}$ ($\approx$0.115$\omega_{pd}$) is used in the simulations. The dust particles are evolved for the 120000 time steps for the equilibration of the system. Fluctuations in temperature, pressure, and total energy of the system over time are monitored to verify system equilibration. After equilibration, the dust particles settle into either a solid (BCC or FCC) or liquid state depending on the values of $\Gamma$ and $\kappa$, see ref.~\cite{Hamaguchi_97}. However, in the solid phase, the random initial distribution does not settle into exact BCC or FCC lattice sites.

\section*{Appendix B: The Formulation of Elastic Stiffness Constants}
\label{sec_app_elas_elas}
\setcounter{equation}{0}
\addcontentsline{toc}{section}{Appendix}
\renewcommand{\theequation}{B\arabic{equation}}
Using Voigt notation, the elastic constants are defined by the following relation~\cite{Nieves_22,Kittel_18}:
\begin{equation}
\sigma_i = \sum_{j=1}^{6} c_{ij} \epsilon_{j}\,, \hspace{1cm}        i=1,......,6
\label{eq:voigt_eq}
\end{equation}
In the above equation, $\sigma_i$, $c_{ij}$, and $\epsilon_j$ are the components of stress tensor, elastic stiffness constants, and elements of strain tensors, respectively. The elastic stiffness constants $c_{ij}$ are calculated using elastic modulus $d_j$ of the crystal along various directions, namely X, Y, Z, XY, XZ, and YZ, which are given by
\begin{equation}
c_{ij} =  [d_j^p + d_j^n]/2\,,
\end{equation}
in which i and j run from 1 to 6, and for a given $i$, $j$ runs from 1 to 6. The superscript $p$ and $n$ represent expansion and compression contribution, respectively.  The elastic modulus ($d_j$) are calculated from the stress to strain ratio as follows
\begin{equation}
d_1= - \frac{P_{xx} - P_{xx0}}{(L_x-L_{x0})/L_{x0}} \, ,
\end{equation}

\begin{equation}
d_2= - \frac{P_{yy} - P_{yy0}}{(L_y-L_{y0})/L_{y0}} \, ,
\end{equation}

\begin{equation}
d_3= - \frac{P_{zz} - P_{zz0}}{(L_z-L_{z0})/L_{z0}} \, ,
\end{equation}

\begin{equation}
d_4= - \frac{P_{yz} - P_{yz0}}{(L_{z}-L_{z0})/L_{z0}} \, ,
\end{equation}

\begin{equation}
d_5= - \frac{P_{xz} - P_{xz0}}{(L_z-L_{z0})/L_{z0}} \, ,
\end{equation}

\begin{equation}
d_6= - \frac{P_{xy} - P_{xy0}}{(L_y-L_{y0})/L_{y0}} \, .
\end{equation}
Here, the number 1 to 6 specifies deformation along X, Y, Z, XY, XZ, and YZ directions, respectively, $P_{\alpha\beta0}$ and $P_{\alpha\beta}$ correspond to initial and final stress components of the crystal, respectively, and $L_{i0}$ and $L_i$ are the initial and final simulation box length along a given direction, respectively. The stress of the Yukawa system reads
\begin{equation}
P_{\alpha\beta} = \frac{1}{V}\left[ \sum_{k}^{N} mv_{k\alpha} v_{k\beta} - \sum_{k}^{N^\prime} r_{k\alpha} \nabla U_{k\beta} \right] \,,
\label{eq:stress_ex}
\end{equation}
where $V$, $N$ and $N^\prime$, $m$, and $v$ are the volume of the simulation cell, the number of particles, the mass of the particle, and the velocity of the particle, respectively. Finally, the elastic stiffness constants used in the calculation of bulk modulus ($K$), shear modulus ($G$), and Poisson's ratio ($\nu$) are calculated as 
\begin{equation}
C_{11} = (c_{11} + c_{22} + c_{33})/3\,,
\end{equation}
\begin{equation}
C_{12} = (c_{12} + c_{13} + c_{23})/3\,,
\end{equation}
\begin{equation}
C_{44} = (c_{44} + c_{55} + c_{66})/3\,.
\label{eq:stress_c44}
\end{equation}
%

\begin{thebibliography}{53}%
\makeatletter
\providecommand \@ifxundefined [1]{%
 \@ifx{#1\undefined}
}%
\providecommand \@ifnum [1]{%
 \ifnum #1\expandafter \@firstoftwo
 \else \expandafter \@secondoftwo
 \fi
}%
\providecommand \@ifx [1]{%
 \ifx #1\expandafter \@firstoftwo
 \else \expandafter \@secondoftwo
 \fi
}%
\providecommand \natexlab [1]{#1}%
\providecommand \enquote  [1]{``#1''}%
\providecommand \bibnamefont  [1]{#1}%
\providecommand \bibfnamefont [1]{#1}%
\providecommand \citenamefont [1]{#1}%
\providecommand \href@noop [0]{\@secondoftwo}%
\providecommand \href [0]{\begingroup \@sanitize@url \@href}%
\providecommand \@href[1]{\@@startlink{#1}\@@href}%
\providecommand \@@href[1]{\endgroup#1\@@endlink}%
\providecommand \@sanitize@url [0]{\catcode `\\12\catcode `\$12\catcode
  `\&12\catcode `\#12\catcode `\^12\catcode `\_12\catcode `\%12\relax}%
\providecommand \@@startlink[1]{}%
\providecommand \@@endlink[0]{}%
\providecommand \url  [0]{\begingroup\@sanitize@url \@url }%
\providecommand \@url [1]{\endgroup\@href {#1}{\urlprefix }}%
\providecommand \urlprefix  [0]{URL }%
\providecommand \Eprint [0]{\href }%
\providecommand \doibase [0]{http://dx.doi.org/}%
\providecommand \selectlanguage [0]{\@gobble}%
\providecommand \bibinfo  [0]{\@secondoftwo}%
\providecommand \bibfield  [0]{\@secondoftwo}%
\providecommand \translation [1]{[#1]}%
\providecommand \BibitemOpen [0]{}%
\providecommand \bibitemStop [0]{}%
\providecommand \bibitemNoStop [0]{.\EOS\space}%
\providecommand \EOS [0]{\spacefactor3000\relax}%
\providecommand \BibitemShut  [1]{\csname bibitem#1\endcsname}%
\let\auto@bib@innerbib\@empty
\bibitem [{\citenamefont {Shukla}\ and\ \citenamefont
  {Mamun}(2015)}]{Shukla_15}%
  \BibitemOpen
  \bibfield  {author} {\bibinfo {author} {\bibfnamefont {P.~K.}\ \bibnamefont
  {Shukla}}\ and\ \bibinfo {author} {\bibfnamefont {A.}~\bibnamefont {Mamun}},\
  }\href@noop {} {\emph {\bibinfo {title} {Introduction to dusty plasma
  physics}}}\ (\bibinfo  {publisher} {CRC press},\ \bibinfo {year}
  {2015})\BibitemShut {NoStop}%
\bibitem [{\citenamefont {Chu}\ and\ \citenamefont {Lin}(1994)}]{Chu_94}%
  \BibitemOpen
  \bibfield  {author} {\bibinfo {author} {\bibfnamefont {J.}~\bibnamefont
  {Chu}}\ and\ \bibinfo {author} {\bibfnamefont {I.}~\bibnamefont {Lin}},\
  }\bibfield  {title} {\enquote {\bibinfo {title} {Direct observation of
  coulomb crystals and liquids in strongly coupled rf dusty plasmas},}\
  }\href@noop {} {\bibfield  {journal} {\bibinfo  {journal} {Phys. Rev. Lett.}\
  }\textbf {\bibinfo {volume} {72}},\ \bibinfo {pages} {4009} (\bibinfo {year}
  {1994})}\BibitemShut {NoStop}%
\bibitem [{\citenamefont {Chaubey}\ and\ \citenamefont
  {Goree}(2022)}]{Chaubey_22}%
  \BibitemOpen
  \bibfield  {author} {\bibinfo {author} {\bibfnamefont {N.}~\bibnamefont
  {Chaubey}}\ and\ \bibinfo {author} {\bibfnamefont {J.}~\bibnamefont
  {Goree}},\ }\bibfield  {title} {\enquote {\bibinfo {title} {Coulomb expansion
  of a thin dust cloud observed experimentally under afterglow plasma
  conditions},}\ }\href@noop {} {\bibfield  {journal} {\bibinfo  {journal}
  {Phys. Plasmas}\ }\textbf {\bibinfo {volume} {29}},\ \bibinfo {pages}
  {113705} (\bibinfo {year} {2022})}\BibitemShut {NoStop}%
\bibitem [{\citenamefont {Morfill}\ and\ \citenamefont
  {Ivlev}(2009)}]{Morfill_09}%
  \BibitemOpen
  \bibfield  {author} {\bibinfo {author} {\bibfnamefont {G.~E.}\ \bibnamefont
  {Morfill}}\ and\ \bibinfo {author} {\bibfnamefont {A.~V.}\ \bibnamefont
  {Ivlev}},\ }\bibfield  {title} {\enquote {\bibinfo {title} {Complex plasmas:
  An interdisciplinary research field},}\ }\href@noop {} {\bibfield  {journal}
  {\bibinfo  {journal} {Rev. Mod. Phys.}\ }\textbf {\bibinfo {volume} {81}},\
  \bibinfo {pages} {1353} (\bibinfo {year} {2009})}\BibitemShut {NoStop}%
\bibitem [{\citenamefont {Thomas}\ \emph {et~al.}(1994)\citenamefont {Thomas},
  \citenamefont {Morfill}, \citenamefont {Demmel}, \citenamefont {Goree},
  \citenamefont {Feuerbacher},\ and\ \citenamefont {M\"ohlmann}}]{Thomas_94}%
  \BibitemOpen
  \bibfield  {author} {\bibinfo {author} {\bibfnamefont {H.}~\bibnamefont
  {Thomas}}, \bibinfo {author} {\bibfnamefont {G.~E.}\ \bibnamefont {Morfill}},
  \bibinfo {author} {\bibfnamefont {V.}~\bibnamefont {Demmel}}, \bibinfo
  {author} {\bibfnamefont {J.}~\bibnamefont {Goree}}, \bibinfo {author}
  {\bibfnamefont {B.}~\bibnamefont {Feuerbacher}}, \ and\ \bibinfo {author}
  {\bibfnamefont {D.}~\bibnamefont {M\"ohlmann}},\ }\bibfield  {title}
  {\enquote {\bibinfo {title} {Plasma crystal: Coulomb crystallization in a
  dusty plasma},}\ }\href@noop {} {\bibfield  {journal} {\bibinfo  {journal}
  {Phys. Rev. Lett.}\ }\textbf {\bibinfo {volume} {73}},\ \bibinfo {pages}
  {652--655} (\bibinfo {year} {1994})}\BibitemShut {NoStop}%
\bibitem [{\citenamefont {Hamaguchi}, \citenamefont {Farouki},\ and\
  \citenamefont {Dubin}(1997)}]{Hamaguchi_97}%
  \BibitemOpen
  \bibfield  {author} {\bibinfo {author} {\bibfnamefont {S.}~\bibnamefont
  {Hamaguchi}}, \bibinfo {author} {\bibfnamefont {R.}~\bibnamefont {Farouki}},
  \ and\ \bibinfo {author} {\bibfnamefont {D.}~\bibnamefont {Dubin}},\
  }\bibfield  {title} {\enquote {\bibinfo {title} {Triple point of yukawa
  systems},}\ }\href@noop {} {\bibfield  {journal} {\bibinfo  {journal} {Phys.
  Rev. E}\ }\textbf {\bibinfo {volume} {56}},\ \bibinfo {pages} {4671}
  (\bibinfo {year} {1997})}\BibitemShut {NoStop}%
\bibitem [{\citenamefont {Maity}\ and\ \citenamefont {Das}(2019)}]{Maity_19}%
  \BibitemOpen
  \bibfield  {author} {\bibinfo {author} {\bibfnamefont {S.}~\bibnamefont
  {Maity}}\ and\ \bibinfo {author} {\bibfnamefont {A.}~\bibnamefont {Das}},\
  }\bibfield  {title} {\enquote {\bibinfo {title} {Molecular dynamics study of
  crystal formation and structural phase transition in yukawa system for dusty
  plasma medium},}\ }\href@noop {} {\bibfield  {journal} {\bibinfo  {journal}
  {Phys. Plasmas}\ }\textbf {\bibinfo {volume} {26}},\ \bibinfo {pages}
  {023703} (\bibinfo {year} {2019})}\BibitemShut {NoStop}%
\bibitem [{\citenamefont {Maity}\ \emph {et~al.}(2018)\citenamefont {Maity},
  \citenamefont {Das}, \citenamefont {Kumar},\ and\ \citenamefont
  {Tiwari}}]{Maity_18}%
  \BibitemOpen
  \bibfield  {author} {\bibinfo {author} {\bibfnamefont {S.}~\bibnamefont
  {Maity}}, \bibinfo {author} {\bibfnamefont {A.}~\bibnamefont {Das}}, \bibinfo
  {author} {\bibfnamefont {S.}~\bibnamefont {Kumar}}, \ and\ \bibinfo {author}
  {\bibfnamefont {S.~K.}\ \bibnamefont {Tiwari}},\ }\bibfield  {title}
  {\enquote {\bibinfo {title} {Interplay of single particle and collective
  response in molecular dynamics simulation of dusty plasma system},}\
  }\href@noop {} {\bibfield  {journal} {\bibinfo  {journal} {Phys. Plasmas}\
  }\textbf {\bibinfo {volume} {25}},\ \bibinfo {pages} {043705} (\bibinfo
  {year} {2018})}\BibitemShut {NoStop}%
\bibitem [{\citenamefont {Deshwal}\ \emph {et~al.}(2022)\citenamefont
  {Deshwal}, \citenamefont {Yadav}, \citenamefont {Prasad}, \citenamefont
  {Sridev}, \citenamefont {Ahuja}, \citenamefont {Maity},\ and\ \citenamefont
  {Das}}]{Deshwal_22}%
  \BibitemOpen
  \bibfield  {author} {\bibinfo {author} {\bibfnamefont {P.}~\bibnamefont
  {Deshwal}}, \bibinfo {author} {\bibfnamefont {M.}~\bibnamefont {Yadav}},
  \bibinfo {author} {\bibfnamefont {C.}~\bibnamefont {Prasad}}, \bibinfo
  {author} {\bibfnamefont {S.}~\bibnamefont {Sridev}}, \bibinfo {author}
  {\bibfnamefont {Y.}~\bibnamefont {Ahuja}}, \bibinfo {author} {\bibfnamefont
  {S.}~\bibnamefont {Maity}}, \ and\ \bibinfo {author} {\bibfnamefont
  {A.}~\bibnamefont {Das}},\ }\bibfield  {title} {\enquote {\bibinfo {title}
  {Chaotic dynamics of small-sized charged yukawa dust clusters},}\ }\href@noop
  {} {\bibfield  {journal} {\bibinfo  {journal} {Chaos: An Interdisciplinary
  Journal of Nonlinear Science}\ }\textbf {\bibinfo {volume} {32}},\ \bibinfo
  {pages} {063136} (\bibinfo {year} {2022})}\BibitemShut {NoStop}%
\bibitem [{\citenamefont {Kumar}, \citenamefont {Tiwari},\ and\ \citenamefont
  {Das}(2017)}]{Sandeep_17}%
  \BibitemOpen
  \bibfield  {author} {\bibinfo {author} {\bibfnamefont {S.}~\bibnamefont
  {Kumar}}, \bibinfo {author} {\bibfnamefont {S.~K.}\ \bibnamefont {Tiwari}}, \
  and\ \bibinfo {author} {\bibfnamefont {A.}~\bibnamefont {Das}},\ }\bibfield
  {title} {\enquote {\bibinfo {title} {Observation of the korteweg-de vries
  soliton in molecular dynamics simulations of a dusty plasma medium},}\
  }\href@noop {} {\bibfield  {journal} {\bibinfo  {journal} {Phys. Plasmas}\
  }\textbf {\bibinfo {volume} {24}},\ \bibinfo {pages} {033711} (\bibinfo
  {year} {2017})}\BibitemShut {NoStop}%
\bibitem [{\citenamefont {Bandyopadhyay}\ \emph {et~al.}(2008)\citenamefont
  {Bandyopadhyay}, \citenamefont {Prasad}, \citenamefont {Sen},\ and\
  \citenamefont {Kaw}}]{Bandyopadhyay_08}%
  \BibitemOpen
  \bibfield  {author} {\bibinfo {author} {\bibfnamefont {P.}~\bibnamefont
  {Bandyopadhyay}}, \bibinfo {author} {\bibfnamefont {G.}~\bibnamefont
  {Prasad}}, \bibinfo {author} {\bibfnamefont {A.}~\bibnamefont {Sen}}, \ and\
  \bibinfo {author} {\bibfnamefont {P.}~\bibnamefont {Kaw}},\ }\bibfield
  {title} {\enquote {\bibinfo {title} {Experimental study of nonlinear dust
  acoustic solitary waves in a dusty plasma},}\ }\href@noop {} {\bibfield
  {journal} {\bibinfo  {journal} {Phys. Rev. Lett.}\ }\textbf {\bibinfo
  {volume} {101}},\ \bibinfo {pages} {065006} (\bibinfo {year}
  {2008})}\BibitemShut {NoStop}%
\bibitem [{\citenamefont {Donk{\'o}}\ \emph {et~al.}(2020)\citenamefont
  {Donk{\'o}}, \citenamefont {Hartmann}, \citenamefont {Masheyeva},\ and\
  \citenamefont {Dzhumagulova}}]{Donko_20}%
  \BibitemOpen
  \bibfield  {author} {\bibinfo {author} {\bibfnamefont {Z.}~\bibnamefont
  {Donk{\'o}}}, \bibinfo {author} {\bibfnamefont {P.}~\bibnamefont {Hartmann}},
  \bibinfo {author} {\bibfnamefont {R.~U.}\ \bibnamefont {Masheyeva}}, \ and\
  \bibinfo {author} {\bibfnamefont {K.~N.}\ \bibnamefont {Dzhumagulova}},\
  }\bibfield  {title} {\enquote {\bibinfo {title} {Molecular dynamics
  investigation of soliton propagation in a two-dimensional yukawa liquid},}\
  }\href@noop {} {\bibfield  {journal} {\bibinfo  {journal} {Contrib. Plasma
  Phys.}\ }\textbf {\bibinfo {volume} {60}},\ \bibinfo {pages} {e201900197}
  (\bibinfo {year} {2020})}\BibitemShut {NoStop}%
\bibitem [{\citenamefont {Kumar}\ and\ \citenamefont
  {Sharma}(2023)}]{Prince_23}%
  \BibitemOpen
  \bibfield  {author} {\bibinfo {author} {\bibfnamefont {P.}~\bibnamefont
  {Kumar}}\ and\ \bibinfo {author} {\bibfnamefont {D.}~\bibnamefont {Sharma}},\
  }\bibfield  {title} {\enquote {\bibinfo {title} {Quasi-localized charge
  approximation approach for the nonlinear structures in strongly coupled
  yukawa systems},}\ }\href@noop {} {\bibfield  {journal} {\bibinfo  {journal}
  {Phys. Plasmas}\ }\textbf {\bibinfo {volume} {30}},\ \bibinfo {pages}
  {033702} (\bibinfo {year} {2023})}\BibitemShut {NoStop}%
\bibitem [{\citenamefont {Sharma}\ \emph {et~al.}(2016)\citenamefont {Sharma},
  \citenamefont {Boruah}, \citenamefont {Nakamura},\ and\ \citenamefont
  {Bailung}}]{Sharma_16}%
  \BibitemOpen
  \bibfield  {author} {\bibinfo {author} {\bibfnamefont {S.~K.}\ \bibnamefont
  {Sharma}}, \bibinfo {author} {\bibfnamefont {A.}~\bibnamefont {Boruah}},
  \bibinfo {author} {\bibfnamefont {Y.}~\bibnamefont {Nakamura}}, \ and\
  \bibinfo {author} {\bibfnamefont {H.}~\bibnamefont {Bailung}},\ }\bibfield
  {title} {\enquote {\bibinfo {title} {Observation of dust acoustic shock wave
  in a strongly coupled dusty plasma},}\ }\href@noop {} {\bibfield  {journal}
  {\bibinfo  {journal} {Phys. Plasmas}\ }\textbf {\bibinfo {volume} {23}},\
  \bibinfo {pages} {053702} (\bibinfo {year} {2016})}\BibitemShut {NoStop}%
\bibitem [{\citenamefont {Arora}\ and\ \citenamefont {Maity}(2023)}]{Arora_23}%
  \BibitemOpen
  \bibfield  {author} {\bibinfo {author} {\bibfnamefont {G.}~\bibnamefont
  {Arora}}\ and\ \bibinfo {author} {\bibfnamefont {S.}~\bibnamefont {Maity}},\
  }\bibfield  {title} {\enquote {\bibinfo {title} {Self-excited converging
  shock structure in a complex plasma medium},}\ }\href@noop {} {\bibfield
  {journal} {\bibinfo  {journal} {Phys. Rev. E}\ }\textbf {\bibinfo {volume}
  {108}},\ \bibinfo {pages} {055209} (\bibinfo {year} {2023})}\BibitemShut
  {NoStop}%
\bibitem [{\citenamefont {Kumar}, \citenamefont {Patel},\ and\ \citenamefont
  {Das}(2018)}]{Sandeep_18_POP}%
  \BibitemOpen
  \bibfield  {author} {\bibinfo {author} {\bibfnamefont {S.}~\bibnamefont
  {Kumar}}, \bibinfo {author} {\bibfnamefont {B.}~\bibnamefont {Patel}}, \ and\
  \bibinfo {author} {\bibfnamefont {A.}~\bibnamefont {Das}},\ }\bibfield
  {title} {\enquote {\bibinfo {title} {Spiral waves in driven dusty plasma
  medium: Generalized hydrodynamic fluid description},}\ }\href@noop {}
  {\bibfield  {journal} {\bibinfo  {journal} {Phys. Plasmas}\ }\textbf
  {\bibinfo {volume} {25}},\ \bibinfo {pages} {043701} (\bibinfo {year}
  {2018})}\BibitemShut {NoStop}%
\bibitem [{\citenamefont {Kumar}\ and\ \citenamefont
  {Das}(2018)}]{Sandeep_18_PRE}%
  \BibitemOpen
  \bibfield  {author} {\bibinfo {author} {\bibfnamefont {S.}~\bibnamefont
  {Kumar}}\ and\ \bibinfo {author} {\bibfnamefont {A.}~\bibnamefont {Das}},\
  }\bibfield  {title} {\enquote {\bibinfo {title} {Spiral waves in driven
  strongly coupled yukawa systems},}\ }\href@noop {} {\bibfield  {journal}
  {\bibinfo  {journal} {Phys. Rev. E}\ }\textbf {\bibinfo {volume} {97}},\
  \bibinfo {pages} {063202} (\bibinfo {year} {2018})}\BibitemShut {NoStop}%
\bibitem [{\citenamefont {Bailung}\ \emph {et~al.}(2020)\citenamefont
  {Bailung}, \citenamefont {Chutia}, \citenamefont {Deka}, \citenamefont
  {Boruah}, \citenamefont {Sharma}, \citenamefont {Kumar}, \citenamefont
  {Chutia}, \citenamefont {Nakamura},\ and\ \citenamefont
  {Bailung}}]{Bailung_20}%
  \BibitemOpen
  \bibfield  {author} {\bibinfo {author} {\bibfnamefont {Y.}~\bibnamefont
  {Bailung}}, \bibinfo {author} {\bibfnamefont {B.}~\bibnamefont {Chutia}},
  \bibinfo {author} {\bibfnamefont {T.}~\bibnamefont {Deka}}, \bibinfo {author}
  {\bibfnamefont {A.}~\bibnamefont {Boruah}}, \bibinfo {author} {\bibfnamefont
  {S.~K.}\ \bibnamefont {Sharma}}, \bibinfo {author} {\bibfnamefont
  {S.}~\bibnamefont {Kumar}}, \bibinfo {author} {\bibfnamefont
  {J.}~\bibnamefont {Chutia}}, \bibinfo {author} {\bibfnamefont
  {Y.}~\bibnamefont {Nakamura}}, \ and\ \bibinfo {author} {\bibfnamefont
  {H.}~\bibnamefont {Bailung}},\ }\bibfield  {title} {\enquote {\bibinfo
  {title} {Vortex formation in a strongly coupled dusty plasma flow past an
  obstacle},}\ }\href@noop {} {\bibfield  {journal} {\bibinfo  {journal} {Phys.
  Plasmas}\ }\textbf {\bibinfo {volume} {27}},\ \bibinfo {pages} {123702}
  (\bibinfo {year} {2020})}\BibitemShut {NoStop}%
\bibitem [{\citenamefont {Gupta}, \citenamefont {Ganesh},\ and\ \citenamefont
  {Joy}(2016)}]{Gupta_16}%
  \BibitemOpen
  \bibfield  {author} {\bibinfo {author} {\bibfnamefont {A.}~\bibnamefont
  {Gupta}}, \bibinfo {author} {\bibfnamefont {R.}~\bibnamefont {Ganesh}}, \
  and\ \bibinfo {author} {\bibfnamefont {A.}~\bibnamefont {Joy}},\ }\bibfield
  {title} {\enquote {\bibinfo {title} {Molecular shear heating and vortex
  dynamics in thermostatted two dimensional yukawa liquids},}\ }\href@noop {}
  {\bibfield  {journal} {\bibinfo  {journal} {Phys. Plasmas}\ }\textbf
  {\bibinfo {volume} {23}},\ \bibinfo {pages} {073706} (\bibinfo {year}
  {2016})}\BibitemShut {NoStop}%
\bibitem [{\citenamefont {Dharodi}(2020)}]{Dharodi_20}%
  \BibitemOpen
  \bibfield  {author} {\bibinfo {author} {\bibfnamefont {V.~S.}\ \bibnamefont
  {Dharodi}},\ }\bibfield  {title} {\enquote {\bibinfo {title} {Rotating
  vortices in two-dimensional inhomogeneous strongly coupled dusty plasmas:
  Shear and spiral density waves},}\ }\href@noop {} {\bibfield  {journal}
  {\bibinfo  {journal} {Phys. Rev. E}\ }\textbf {\bibinfo {volume} {102}},\
  \bibinfo {pages} {043216} (\bibinfo {year} {2020})}\BibitemShut {NoStop}%
\bibitem [{\citenamefont {Samsonov}\ \emph {et~al.}(1999)\citenamefont
  {Samsonov}, \citenamefont {Goree}, \citenamefont {Ma}, \citenamefont
  {Bhattacharjee}, \citenamefont {Thomas},\ and\ \citenamefont
  {Morfill}}]{Samsonov_99}%
  \BibitemOpen
  \bibfield  {author} {\bibinfo {author} {\bibfnamefont {D.}~\bibnamefont
  {Samsonov}}, \bibinfo {author} {\bibfnamefont {J.}~\bibnamefont {Goree}},
  \bibinfo {author} {\bibfnamefont {Z.}~\bibnamefont {Ma}}, \bibinfo {author}
  {\bibfnamefont {A.}~\bibnamefont {Bhattacharjee}}, \bibinfo {author}
  {\bibfnamefont {H.}~\bibnamefont {Thomas}}, \ and\ \bibinfo {author}
  {\bibfnamefont {G.}~\bibnamefont {Morfill}},\ }\bibfield  {title} {\enquote
  {\bibinfo {title} {Mach cones in a coulomb lattice and a dusty plasma},}\
  }\href@noop {} {\bibfield  {journal} {\bibinfo  {journal} {Phys. Rev. Lett.}\
  }\textbf {\bibinfo {volume} {83}},\ \bibinfo {pages} {3649} (\bibinfo {year}
  {1999})}\BibitemShut {NoStop}%
\bibitem [{\citenamefont {Wani}, \citenamefont {Verma},\ and\ \citenamefont
  {Tiwari}(2024)}]{Wani_24}%
  \BibitemOpen
  \bibfield  {author} {\bibinfo {author} {\bibfnamefont {R.}~\bibnamefont
  {Wani}}, \bibinfo {author} {\bibfnamefont {M.}~\bibnamefont {Verma}}, \ and\
  \bibinfo {author} {\bibfnamefont {S.}~\bibnamefont {Tiwari}},\ }\bibfield
  {title} {\enquote {\bibinfo {title} {Rayleigh–taylor turbulence in strongly
  coupled dusty plasmas},}\ }\href@noop {} {\bibfield  {journal} {\bibinfo
  {journal} {Phys. Plasmas}\ }\textbf {\bibinfo {volume} {31}},\ \bibinfo
  {pages} {082306} (\bibinfo {year} {2024})}\BibitemShut {NoStop}%
\bibitem [{\citenamefont {Wolter}\ and\ \citenamefont
  {Melzer}(2005)}]{Wolter_05}%
  \BibitemOpen
  \bibfield  {author} {\bibinfo {author} {\bibfnamefont {M.}~\bibnamefont
  {Wolter}}\ and\ \bibinfo {author} {\bibfnamefont {A.}~\bibnamefont
  {Melzer}},\ }\bibfield  {title} {\enquote {\bibinfo {title} {Laser heating of
  particles in dusty plasmas},}\ }\href@noop {} {\bibfield  {journal} {\bibinfo
   {journal} {Phys. Rev. E—Statistical, Nonlinear, and Soft Matter Physics}\
  }\textbf {\bibinfo {volume} {71}},\ \bibinfo {pages} {036414} (\bibinfo
  {year} {2005})}\BibitemShut {NoStop}%
\bibitem [{\citenamefont {Nosenko}\ and\ \citenamefont
  {Goree}(2004)}]{Nosenko_04}%
  \BibitemOpen
  \bibfield  {author} {\bibinfo {author} {\bibfnamefont {V.}~\bibnamefont
  {Nosenko}}\ and\ \bibinfo {author} {\bibfnamefont {J.}~\bibnamefont
  {Goree}},\ }\bibfield  {title} {\enquote {\bibinfo {title} {Shear flows and
  shear viscosity in a two-dimensional yukawa system (dusty plasma)},}\
  }\href@noop {} {\bibfield  {journal} {\bibinfo  {journal} {Phys. Rev. Lett.}\
  }\textbf {\bibinfo {volume} {93}},\ \bibinfo {pages} {155004} (\bibinfo
  {year} {2004})}\BibitemShut {NoStop}%
\bibitem [{\citenamefont {Konopka}, \citenamefont {Morfill},\ and\
  \citenamefont {Ratke}(2000)}]{Konopka_00}%
  \BibitemOpen
  \bibfield  {author} {\bibinfo {author} {\bibfnamefont {U.}~\bibnamefont
  {Konopka}}, \bibinfo {author} {\bibfnamefont {G.~E.}\ \bibnamefont
  {Morfill}}, \ and\ \bibinfo {author} {\bibfnamefont {L.}~\bibnamefont
  {Ratke}},\ }\bibfield  {title} {\enquote {\bibinfo {title} {Measurement of
  the interaction potential of microspheres in the sheath of a rf discharge},}\
  }\href@noop {} {\bibfield  {journal} {\bibinfo  {journal} {Phys. Rev. Lett.}\
  }\textbf {\bibinfo {volume} {84}},\ \bibinfo {pages} {891--894} (\bibinfo
  {year} {2000})}\BibitemShut {NoStop}%
\bibitem [{\citenamefont {Palberg}\ \emph {et~al.}(1995)\citenamefont
  {Palberg}, \citenamefont {M\"onch}, \citenamefont {Bitzer}, \citenamefont
  {Piazza},\ and\ \citenamefont {Bellini}}]{Palberg}%
  \BibitemOpen
  \bibfield  {author} {\bibinfo {author} {\bibfnamefont {T.}~\bibnamefont
  {Palberg}}, \bibinfo {author} {\bibfnamefont {W.}~\bibnamefont {M\"onch}},
  \bibinfo {author} {\bibfnamefont {F.}~\bibnamefont {Bitzer}}, \bibinfo
  {author} {\bibfnamefont {R.}~\bibnamefont {Piazza}}, \ and\ \bibinfo {author}
  {\bibfnamefont {T.}~\bibnamefont {Bellini}},\ }\bibfield  {title} {\enquote
  {\bibinfo {title} {Freezing transition for colloids with adjustable charge: A
  test of charge renormalization},}\ }\href@noop {} {\bibfield  {journal}
  {\bibinfo  {journal} {Phys. Rev. Lett.}\ }\textbf {\bibinfo {volume} {74}},\
  \bibinfo {pages} {4555--4558} (\bibinfo {year} {1995})}\BibitemShut {NoStop}%
\bibitem [{\citenamefont {Terao}\ and\ \citenamefont {Nakayama}(1999)}]{Terao}%
  \BibitemOpen
  \bibfield  {author} {\bibinfo {author} {\bibfnamefont {T.}~\bibnamefont
  {Terao}}\ and\ \bibinfo {author} {\bibfnamefont {T.}~\bibnamefont
  {Nakayama}},\ }\bibfield  {title} {\enquote {\bibinfo {title}
  {Crystallization in quasi-two-dimensional colloidal systems at an air-water
  interface},}\ }\href@noop {} {\bibfield  {journal} {\bibinfo  {journal}
  {Phys. Rev. E}\ }\textbf {\bibinfo {volume} {60}},\ \bibinfo {pages}
  {7157--7162} (\bibinfo {year} {1999})}\BibitemShut {NoStop}%
\bibitem [{\citenamefont {Levin}(2002)}]{Levin}%
  \BibitemOpen
  \bibfield  {author} {\bibinfo {author} {\bibfnamefont {Y.}~\bibnamefont
  {Levin}},\ }\bibfield  {title} {\enquote {\bibinfo {title} {Electrostatic
  correlations: from plasma to biology},}\ }\href@noop {} {\bibfield  {journal}
  {\bibinfo  {journal} {Rep. Prog. Phys.}\ }\textbf {\bibinfo {volume} {65}},\
  \bibinfo {pages} {1577} (\bibinfo {year} {2002})}\BibitemShut {NoStop}%
\bibitem [{\citenamefont {Lee}\ \emph {et~al.}(2017)\citenamefont {Lee},
  \citenamefont {Perez-Martinez}, \citenamefont {Smith},\ and\ \citenamefont
  {Perkin}}]{Lee}%
  \BibitemOpen
  \bibfield  {author} {\bibinfo {author} {\bibfnamefont {A.~A.}\ \bibnamefont
  {Lee}}, \bibinfo {author} {\bibfnamefont {C.~S.}\ \bibnamefont
  {Perez-Martinez}}, \bibinfo {author} {\bibfnamefont {A.~M.}\ \bibnamefont
  {Smith}}, \ and\ \bibinfo {author} {\bibfnamefont {S.}~\bibnamefont
  {Perkin}},\ }\bibfield  {title} {\enquote {\bibinfo {title} {Scaling analysis
  of the screening length in concentrated electrolytes},}\ }\href@noop {}
  {\bibfield  {journal} {\bibinfo  {journal} {Phys. Rev. Lett.}\ }\textbf
  {\bibinfo {volume} {119}},\ \bibinfo {pages} {026002} (\bibinfo {year}
  {2017})}\BibitemShut {NoStop}%
\bibitem [{\citenamefont {Lyon}, \citenamefont {Bergeson},\ and\ \citenamefont
  {Murillo}(2013)}]{Lyon}%
  \BibitemOpen
  \bibfield  {author} {\bibinfo {author} {\bibfnamefont {M.}~\bibnamefont
  {Lyon}}, \bibinfo {author} {\bibfnamefont {S.~D.}\ \bibnamefont {Bergeson}},
  \ and\ \bibinfo {author} {\bibfnamefont {M.~S.}\ \bibnamefont {Murillo}},\
  }\bibfield  {title} {\enquote {\bibinfo {title} {Limit of strong ion coupling
  due to electron shielding},}\ }\href@noop {} {\bibfield  {journal} {\bibinfo
  {journal} {Phys. Rev. E}\ }\textbf {\bibinfo {volume} {87}},\ \bibinfo
  {pages} {033101} (\bibinfo {year} {2013})}\BibitemShut {NoStop}%
\bibitem [{\citenamefont {Vorberger}\ \emph {et~al.}(2012)\citenamefont
  {Vorberger}, \citenamefont {Donko}, \citenamefont {Tkachenko},\ and\
  \citenamefont {Gericke}}]{Vorberger_12}%
  \BibitemOpen
  \bibfield  {author} {\bibinfo {author} {\bibfnamefont {J.}~\bibnamefont
  {Vorberger}}, \bibinfo {author} {\bibfnamefont {Z.}~\bibnamefont {Donko}},
  \bibinfo {author} {\bibfnamefont {I.}~\bibnamefont {Tkachenko}}, \ and\
  \bibinfo {author} {\bibfnamefont {D.~O.}\ \bibnamefont {Gericke}},\
  }\bibfield  {title} {\enquote {\bibinfo {title} {Dynamic ion structure factor
  of warm dense matter},}\ }\href@noop {} {\bibfield  {journal} {\bibinfo
  {journal} {Phys. Rev. Lett.}\ }\textbf {\bibinfo {volume} {109}},\ \bibinfo
  {pages} {225001} (\bibinfo {year} {2012})}\BibitemShut {NoStop}%
\bibitem [{\citenamefont {Liu}\ and\ \citenamefont {Goree}(2017)}]{Liu_17}%
  \BibitemOpen
  \bibfield  {author} {\bibinfo {author} {\bibfnamefont {B.}~\bibnamefont
  {Liu}}\ and\ \bibinfo {author} {\bibfnamefont {J.}~\bibnamefont {Goree}},\
  }\bibfield  {title} {\enquote {\bibinfo {title} {Determination of yield
  stress of 2d (yukawa) dusty plasma},}\ }\href@noop {} {\bibfield  {journal}
  {\bibinfo  {journal} {Phys. Plasmas}\ }\textbf {\bibinfo {volume} {24}},\
  \bibinfo {pages} {103702} (\bibinfo {year} {2017})}\BibitemShut {NoStop}%
\bibitem [{\citenamefont {Ashwin}\ and\ \citenamefont {Sen}(2015)}]{Joy_15}%
  \BibitemOpen
  \bibfield  {author} {\bibinfo {author} {\bibfnamefont {J.}~\bibnamefont
  {Ashwin}}\ and\ \bibinfo {author} {\bibfnamefont {A.}~\bibnamefont {Sen}},\
  }\bibfield  {title} {\enquote {\bibinfo {title} {Microscopic origin of shear
  relaxation in a model viscoelastic liquid},}\ }\href@noop {} {\bibfield
  {journal} {\bibinfo  {journal} {Phys. Rev. Lett.}\ }\textbf {\bibinfo
  {volume} {114}},\ \bibinfo {pages} {055002} (\bibinfo {year}
  {2015})}\BibitemShut {NoStop}%
\bibitem [{\citenamefont {Wang}, \citenamefont {Huang},\ and\ \citenamefont
  {Feng}(2019)}]{Wang_19}%
  \BibitemOpen
  \bibfield  {author} {\bibinfo {author} {\bibfnamefont {K.}~\bibnamefont
  {Wang}}, \bibinfo {author} {\bibfnamefont {D.}~\bibnamefont {Huang}}, \ and\
  \bibinfo {author} {\bibfnamefont {Y.}~\bibnamefont {Feng}},\ }\bibfield
  {title} {\enquote {\bibinfo {title} {Shear modulus of two-dimensional yukawa
  or dusty-plasma solids obtained from the viscoelasticity in the liquid
  state},}\ }\href@noop {} {\bibfield  {journal} {\bibinfo  {journal} {Phys.
  Rev. E}\ }\textbf {\bibinfo {volume} {99}},\ \bibinfo {pages} {063206}
  (\bibinfo {year} {2019})}\BibitemShut {NoStop}%
\bibitem [{\citenamefont {Robbins}, \citenamefont {Kremer},\ and\ \citenamefont
  {Grest}(1988)}]{Robbins_88}%
  \BibitemOpen
  \bibfield  {author} {\bibinfo {author} {\bibfnamefont {M.~O.}\ \bibnamefont
  {Robbins}}, \bibinfo {author} {\bibfnamefont {K.}~\bibnamefont {Kremer}}, \
  and\ \bibinfo {author} {\bibfnamefont {G.~S.}\ \bibnamefont {Grest}},\
  }\bibfield  {title} {\enquote {\bibinfo {title} {Phase diagram and dynamics
  of yukawa systems},}\ }\href@noop {} {\bibfield  {journal} {\bibinfo
  {journal} {J. Chem. Phys.}\ }\textbf {\bibinfo {volume} {88}},\ \bibinfo
  {pages} {3286--3312} (\bibinfo {year} {1988})}\BibitemShut {NoStop}%
\bibitem [{\citenamefont {Khrapak}\ and\ \citenamefont
  {Klumov}(2020)}]{Khrapak_20}%
  \BibitemOpen
  \bibfield  {author} {\bibinfo {author} {\bibfnamefont {S.~A.}\ \bibnamefont
  {Khrapak}}\ and\ \bibinfo {author} {\bibfnamefont {B.~A.}\ \bibnamefont
  {Klumov}},\ }\bibfield  {title} {\enquote {\bibinfo {title} {Instantaneous
  shear modulus of yukawa fluids across coupling regimes},}\ }\href@noop {}
  {\bibfield  {journal} {\bibinfo  {journal} {Phys. Plasmas}\ }\textbf
  {\bibinfo {volume} {27}},\ \bibinfo {pages} {024501} (\bibinfo {year}
  {2020})}\BibitemShut {NoStop}%
\bibitem [{\citenamefont {Kozhberov}(2022)}]{Kozhberov_22}%
  \BibitemOpen
  \bibfield  {author} {\bibinfo {author} {\bibfnamefont {A.~A.}\ \bibnamefont
  {Kozhberov}},\ }\bibfield  {title} {\enquote {\bibinfo {title} {Elastic
  properties of yukawa crystals},}\ }\href@noop {} {\bibfield  {journal}
  {\bibinfo  {journal} {Phys. Plasmas}\ }\textbf {\bibinfo {volume} {29}},\
  \bibinfo {pages} {043701} (\bibinfo {year} {2022})}\BibitemShut {NoStop}%
\bibitem [{\citenamefont {Kittel}\ and\ \citenamefont
  {McEuen}(2018)}]{Kittel_18}%
  \BibitemOpen
  \bibfield  {author} {\bibinfo {author} {\bibfnamefont {C.}~\bibnamefont
  {Kittel}}\ and\ \bibinfo {author} {\bibfnamefont {P.}~\bibnamefont
  {McEuen}},\ }\href@noop {} {\emph {\bibinfo {title} {Introduction to solid
  state physics}}}\ (\bibinfo  {publisher} {John Wiley \& Sons},\ \bibinfo
  {year} {2018})\BibitemShut {NoStop}%
\bibitem [{\citenamefont {Simmons}(1965)}]{Simmons_65}%
  \BibitemOpen
  \bibfield  {author} {\bibinfo {author} {\bibfnamefont {G.}~\bibnamefont
  {Simmons}},\ }\bibfield  {title} {\enquote {\bibinfo {title} {Single crystal
  elastic constants and calculated aggregate properties},}\ }\href@noop {}
  {\bibfield  {journal} {\bibinfo  {journal} {J. Grad. Res.}\ }\textbf
  {\bibinfo {volume} {34}},\ \bibinfo {pages} {1} (\bibinfo {year}
  {1965})}\BibitemShut {NoStop}%
\bibitem [{\citenamefont {Martorell}\ \emph {et~al.}(2013)\citenamefont
  {Martorell}, \citenamefont {Vo{\v{c}}adlo}, \citenamefont {Brodholt},\ and\
  \citenamefont {Wood}}]{Martorell_13}%
  \BibitemOpen
  \bibfield  {author} {\bibinfo {author} {\bibfnamefont {B.}~\bibnamefont
  {Martorell}}, \bibinfo {author} {\bibfnamefont {L.}~\bibnamefont
  {Vo{\v{c}}adlo}}, \bibinfo {author} {\bibfnamefont {J.}~\bibnamefont
  {Brodholt}}, \ and\ \bibinfo {author} {\bibfnamefont {I.~G.}\ \bibnamefont
  {Wood}},\ }\bibfield  {title} {\enquote {\bibinfo {title} {Strong premelting
  effect in the elastic properties of hcp-fe under inner-core conditions},}\
  }\href@noop {} {\bibfield  {journal} {\bibinfo  {journal} {Science}\ }\textbf
  {\bibinfo {volume} {342}},\ \bibinfo {pages} {466--468} (\bibinfo {year}
  {2013})}\BibitemShut {NoStop}%
\bibitem [{\citenamefont {Donk\'o}\ and\ \citenamefont
  {Hartmann}(2008)}]{Donko_08_vis}%
  \BibitemOpen
  \bibfield  {author} {\bibinfo {author} {\bibfnamefont {Z.}~\bibnamefont
  {Donk\'o}}\ and\ \bibinfo {author} {\bibfnamefont {P.}~\bibnamefont
  {Hartmann}},\ }\bibfield  {title} {\enquote {\bibinfo {title} {Shear
  viscosity of strongly coupled yukawa liquids},}\ }\href@noop {} {\bibfield
  {journal} {\bibinfo  {journal} {Phys. Rev. E}\ }\textbf {\bibinfo {volume}
  {78}},\ \bibinfo {pages} {026408} (\bibinfo {year} {2008})}\BibitemShut
  {NoStop}%
\bibitem [{\citenamefont {Morris}\ \emph {et~al.}(1994)\citenamefont {Morris},
  \citenamefont {Wang}, \citenamefont {Ho},\ and\ \citenamefont
  {Chan}}]{Morris_94}%
  \BibitemOpen
  \bibfield  {author} {\bibinfo {author} {\bibfnamefont {J.~R.}\ \bibnamefont
  {Morris}}, \bibinfo {author} {\bibfnamefont {C.}~\bibnamefont {Wang}},
  \bibinfo {author} {\bibfnamefont {K.}~\bibnamefont {Ho}}, \ and\ \bibinfo
  {author} {\bibfnamefont {C.~T.}\ \bibnamefont {Chan}},\ }\bibfield  {title}
  {\enquote {\bibinfo {title} {Melting line of aluminum from simulations of
  coexisting phases},}\ }\href@noop {} {\bibfield  {journal} {\bibinfo
  {journal} {Phys. Rev. B}\ }\textbf {\bibinfo {volume} {49}},\ \bibinfo
  {pages} {3109} (\bibinfo {year} {1994})}\BibitemShut {NoStop}%
\bibitem [{\citenamefont {Zhang}\ and\ \citenamefont
  {Maginn}(2012)}]{Zhang_12}%
  \BibitemOpen
  \bibfield  {author} {\bibinfo {author} {\bibfnamefont {Y.}~\bibnamefont
  {Zhang}}\ and\ \bibinfo {author} {\bibfnamefont {E.~J.}\ \bibnamefont
  {Maginn}},\ }\bibfield  {title} {\enquote {\bibinfo {title} {A comparison of
  methods for melting point calculation using molecular dynamics
  simulations},}\ }\href@noop {} {\bibfield  {journal} {\bibinfo  {journal} {J.
  Chem. Phys.}\ }\textbf {\bibinfo {volume} {136}} (\bibinfo {year}
  {2012})}\BibitemShut {NoStop}%
\bibitem [{\citenamefont {Archer}\ \emph {et~al.}(1972)\citenamefont {Archer},
  \citenamefont {Crandall}, \citenamefont {Dahl},\ and\ \citenamefont
  {Lardner}}]{Archer_72}%
  \BibitemOpen
  \bibfield  {author} {\bibinfo {author} {\bibfnamefont {R.~R.}\ \bibnamefont
  {Archer}}, \bibinfo {author} {\bibfnamefont {S.~H.}\ \bibnamefont
  {Crandall}}, \bibinfo {author} {\bibfnamefont {N.~C.}\ \bibnamefont {Dahl}},
  \ and\ \bibinfo {author} {\bibfnamefont {T.~J.}\ \bibnamefont {Lardner}},\
  }\href@noop {} {\emph {\bibinfo {title} {An introduction to the mechanics of
  solids}}},\ \bibinfo {edition} {2nd}\ ed.\ (\bibinfo  {publisher}
  {McGraw-Hill Kogakusha},\ \bibinfo {year} {1972})\BibitemShut {NoStop}%
\bibitem [{\citenamefont {Mott}\ and\ \citenamefont {Roland}(2009)}]{Mott_09}%
  \BibitemOpen
  \bibfield  {author} {\bibinfo {author} {\bibfnamefont {P.}~\bibnamefont
  {Mott}}\ and\ \bibinfo {author} {\bibfnamefont {C.}~\bibnamefont {Roland}},\
  }\bibfield  {title} {\enquote {\bibinfo {title} {Limits to poisson’s ratio
  in isotropic materials},}\ }\href@noop {} {\bibfield  {journal} {\bibinfo
  {journal} {Phys. Rev. B}\ }\textbf {\bibinfo {volume} {80}},\ \bibinfo
  {pages} {132104} (\bibinfo {year} {2009})}\BibitemShut {NoStop}%
\bibitem [{\citenamefont {Jastrzebski}(1976)}]{Jastrzebski_76}%
  \BibitemOpen
  \bibfield  {author} {\bibinfo {author} {\bibfnamefont {Z.~D.}\ \bibnamefont
  {Jastrzebski}},\ }\href@noop {} {\emph {\bibinfo {title} {Nature and
  properties of engineering materials}}}\ (\bibinfo  {publisher} {John Wiley
  and Sons, Inc., New York},\ \bibinfo {year} {1976})\BibitemShut {NoStop}%
\bibitem [{\citenamefont {Donk{\'o}}, \citenamefont {Kalman},\ and\
  \citenamefont {Hartmann}(2008)}]{Donko_08}%
  \BibitemOpen
  \bibfield  {author} {\bibinfo {author} {\bibfnamefont {Z.}~\bibnamefont
  {Donk{\'o}}}, \bibinfo {author} {\bibfnamefont {G.~J.}\ \bibnamefont
  {Kalman}}, \ and\ \bibinfo {author} {\bibfnamefont {P.}~\bibnamefont
  {Hartmann}},\ }\bibfield  {title} {\enquote {\bibinfo {title} {Dynamical
  correlations and collective excitations of yukawa liquids},}\ }\href@noop {}
  {\bibfield  {journal} {\bibinfo  {journal} {J. Phys. Condens. Matter}\
  }\textbf {\bibinfo {volume} {20}},\ \bibinfo {pages} {413101} (\bibinfo
  {year} {2008})}\BibitemShut {NoStop}%
\bibitem [{\citenamefont {Kumar}\ \emph {et~al.}(2023)\citenamefont {Kumar},
  \citenamefont {Tahmasbi}, \citenamefont {Ramakrishna}, \citenamefont
  {Lokamani}, \citenamefont {Nikolov}, \citenamefont {Tranchida}, \citenamefont
  {Wood},\ and\ \citenamefont {Cangi}}]{Sandeep_23}%
  \BibitemOpen
  \bibfield  {author} {\bibinfo {author} {\bibfnamefont {S.}~\bibnamefont
  {Kumar}}, \bibinfo {author} {\bibfnamefont {H.}~\bibnamefont {Tahmasbi}},
  \bibinfo {author} {\bibfnamefont {K.}~\bibnamefont {Ramakrishna}}, \bibinfo
  {author} {\bibfnamefont {M.}~\bibnamefont {Lokamani}}, \bibinfo {author}
  {\bibfnamefont {S.}~\bibnamefont {Nikolov}}, \bibinfo {author} {\bibfnamefont
  {J.}~\bibnamefont {Tranchida}}, \bibinfo {author} {\bibfnamefont {M.~A.}\
  \bibnamefont {Wood}}, \ and\ \bibinfo {author} {\bibfnamefont
  {A.}~\bibnamefont {Cangi}},\ }\bibfield  {title} {\enquote {\bibinfo {title}
  {Transferable interatomic potential for aluminum from ambient conditions to
  warm dense matter},}\ }\href@noop {} {\bibfield  {journal} {\bibinfo
  {journal} {Phys. Rev. Res.}\ }\textbf {\bibinfo {volume} {5}},\ \bibinfo
  {pages} {033162} (\bibinfo {year} {2023})}\BibitemShut {NoStop}%
\bibitem [{\citenamefont {Thompson}\ \emph {et~al.}(2022)\citenamefont
  {Thompson}, \citenamefont {Aktulga}, \citenamefont {Berger}, \citenamefont
  {Bolintineanu}, \citenamefont {Brown}, \citenamefont {Crozier}, \citenamefont
  {In't~Veld}, \citenamefont {Kohlmeyer}, \citenamefont {Moore}, \citenamefont
  {Nguyen} \emph {et~al.}}]{Thompson_22}%
  \BibitemOpen
  \bibfield  {author} {\bibinfo {author} {\bibfnamefont {A.~P.}\ \bibnamefont
  {Thompson}}, \bibinfo {author} {\bibfnamefont {H.~M.}\ \bibnamefont
  {Aktulga}}, \bibinfo {author} {\bibfnamefont {R.}~\bibnamefont {Berger}},
  \bibinfo {author} {\bibfnamefont {D.~S.}\ \bibnamefont {Bolintineanu}},
  \bibinfo {author} {\bibfnamefont {W.~M.}\ \bibnamefont {Brown}}, \bibinfo
  {author} {\bibfnamefont {P.~S.}\ \bibnamefont {Crozier}}, \bibinfo {author}
  {\bibfnamefont {P.~J.}\ \bibnamefont {In't~Veld}}, \bibinfo {author}
  {\bibfnamefont {A.}~\bibnamefont {Kohlmeyer}}, \bibinfo {author}
  {\bibfnamefont {S.~G.}\ \bibnamefont {Moore}}, \bibinfo {author}
  {\bibfnamefont {T.~D.}\ \bibnamefont {Nguyen}},  \emph {et~al.},\ }\bibfield
  {title} {\enquote {\bibinfo {title} {Lammps-a flexible simulation tool for
  particle-based materials modeling at the atomic, meso, and continuum
  scales},}\ }\href@noop {} {\bibfield  {journal} {\bibinfo  {journal} {Comput.
  Phys. Commun.}\ }\textbf {\bibinfo {volume} {271}},\ \bibinfo {pages}
  {108171} (\bibinfo {year} {2022})}\BibitemShut {NoStop}%
\bibitem [{\citenamefont {Schneider}\ and\ \citenamefont
  {Stoll}(1978)}]{Schneider_78}%
  \BibitemOpen
  \bibfield  {author} {\bibinfo {author} {\bibfnamefont {T.}~\bibnamefont
  {Schneider}}\ and\ \bibinfo {author} {\bibfnamefont {E.}~\bibnamefont
  {Stoll}},\ }\bibfield  {title} {\enquote {\bibinfo {title}
  {Molecular-dynamics study of a three-dimensional one-component model for
  distortive phase transitions},}\ }\href@noop {} {\bibfield  {journal}
  {\bibinfo  {journal} {Phys. Rev. B}\ }\textbf {\bibinfo {volume} {17}},\
  \bibinfo {pages} {1302--1322} (\bibinfo {year} {1978})}\BibitemShut {NoStop}%
\bibitem [{\citenamefont {Plimpton}(1995)}]{Plimpton19951}%
  \BibitemOpen
  \bibfield  {author} {\bibinfo {author} {\bibfnamefont {S.}~\bibnamefont
  {Plimpton}},\ }\bibfield  {title} {\enquote {\bibinfo {title} {Fast parallel
  algorithms for short-range molecular dynamics},}\ }\href@noop {} {\bibfield
  {journal} {\bibinfo  {journal} {J. Comput. Phys.}\ }\textbf {\bibinfo
  {volume} {117}},\ \bibinfo {pages} {1 -- 19} (\bibinfo {year}
  {1995})}\BibitemShut {NoStop}%
\bibitem [{\citenamefont {Schwabe}\ and\ \citenamefont
  {Graves}(2013)}]{Schwabe_2013}%
  \BibitemOpen
  \bibfield  {author} {\bibinfo {author} {\bibfnamefont {M.}~\bibnamefont
  {Schwabe}}\ and\ \bibinfo {author} {\bibfnamefont {D.~B.}\ \bibnamefont
  {Graves}},\ }\bibfield  {title} {\enquote {\bibinfo {title} {Simulating the
  dynamics of complex plasmas},}\ }\href@noop {} {\bibfield  {journal}
  {\bibinfo  {journal} {Phys. Rev. E}\ }\textbf {\bibinfo {volume} {88}},\
  \bibinfo {pages} {023101} (\bibinfo {year} {2013})}\BibitemShut {NoStop}%
\bibitem [{\citenamefont {Nieves}\ \emph {et~al.}(2022)\citenamefont {Nieves},
  \citenamefont {Tranchida}, \citenamefont {Nikolov}, \citenamefont {Fraile},\
  and\ \citenamefont {Legut}}]{Nieves_22}%
  \BibitemOpen
  \bibfield  {author} {\bibinfo {author} {\bibfnamefont {P.}~\bibnamefont
  {Nieves}}, \bibinfo {author} {\bibfnamefont {J.}~\bibnamefont {Tranchida}},
  \bibinfo {author} {\bibfnamefont {S.}~\bibnamefont {Nikolov}}, \bibinfo
  {author} {\bibfnamefont {A.}~\bibnamefont {Fraile}}, \ and\ \bibinfo {author}
  {\bibfnamefont {D.}~\bibnamefont {Legut}},\ }\bibfield  {title} {\enquote
  {\bibinfo {title} {Atomistic simulations of magnetoelastic effects on sound
  velocity},}\ }\href@noop {} {\bibfield  {journal} {\bibinfo  {journal} {Phys.
  Rev. B}\ }\textbf {\bibinfo {volume} {105}},\ \bibinfo {pages} {134430}
  (\bibinfo {year} {2022})}\BibitemShut {NoStop}%
\end{thebibliography}
\end{document}